\documentclass[11pt]{article}

\usepackage[preprint]{acl}

\usepackage{times}
\usepackage{latexsym}

\usepackage{amsfonts}
\usepackage{amsmath}
\usepackage{amssymb}

\usepackage{booktabs}
\usepackage{tabularx}
\usepackage{longtable}
\usepackage{xltabular}
\usepackage{dashrule}

\usepackage{float}
\usepackage{placeins}

\usepackage[table]{xcolor}

\usepackage{tcolorbox}

\usepackage{comment}

\tcbset{
  prompt/.style={
    colback=gray!6,
    colframe=gray!55,
    boxrule=0.6pt,
    arc=2pt,
    left=10pt, right=10pt,
    top=6pt, bottom=6pt,
    toptitle=4pt, bottomtitle=4pt,
    fonttitle=\bfseries\sffamily\small,
    coltitle=white,
    colbacktitle=black!65,
  }
}

\usepackage[T1]{fontenc}

\usepackage[utf8]{inputenc}

\usepackage{microtype}

\usepackage{inconsolata}

\usepackage{graphicx}

\usepackage{tikz}
\usepackage{xcolor}
\usepackage{colortbl}
\usepackage{caption}
\usepackage{subcaption}

\usepackage{xspace}

\newcommand{\lc}[1]{\colorbox{gray!12}{\parbox[t]{\dimexpr0.5\textwidth-2.8cm-2\fboxsep\relax}{\ttfamily\footnotesize #1\strut}}}

\newcommand{\bx}[2]{\colorbox{#1}{\small #2}} 

\title{TheoremGraph: Bridging Formal and Informal Mathematics}

\author{
  \textbf{Simon Kurgan\textsuperscript{*~1}} \\
  \texttt{simku22@uw.edu}
  \And
  \textbf{Evan Wang\textsuperscript{* 1}} \\
  \texttt{aurasoph@uw.edu}
  \And
  \textbf{Eric Leonen\textsuperscript{* 1}} \\
  \texttt{eol05@uw.edu}
  \AND
  \textbf{Sophie Szeto\textsuperscript{1}} \\
  \texttt{slszeto@uw.edu}
  \And
  \textbf{Luke Alexander\textsuperscript{1}} \\
  \texttt{lukealex@uw.edu}
  \And
  \textbf{Artemii Remizov\textsuperscript{1}} \\
  \texttt{aremizov@uw.edu}
  \AND
  \textbf{Jarod Alper\textsuperscript{1}} \\
  \texttt{jarod@uw.edu}
  \And
  \textbf{Giovanni Inchiostro\textsuperscript{1}} \\
  \texttt{ginchios@uw.edu}
  \And
  \textbf{Vasily Ilin\textsuperscript{1}} \\
  \texttt{vilin@uw.edu}
  \\
  \\
  \textsuperscript{1}University of Washington Math AI Lab
  \\
  \textsuperscript{*} Equal contribution
  \\
  \small{
    \textbf{Correspondence:}
    \href{mailto:vilin@uw.edu}{vilin@uw.edu}
  }
}

\begin{document}
\maketitle
\begin{abstract}
Mathematical knowledge is organized around statements and their dependencies, but
this structure is exposed unevenly: informal papers cite mostly at the document
level, while formal libraries record fine-grained dependencies over a much smaller
body of mathematics. We introduce \textbf{TheoremGraph}, a unified statement-level
dependency graph spanning both informal and formal mathematics. On the informal side,
we parse $11.7$M theorem-like environments from mathematics arXiv and recover $18.3$M
candidate directed dependencies, each labeled by the extractor that proposed it so
downstream users can trade coverage for precision. On the formal side, we release
\textbf{LeanGraph}, a Lean~4 elaborator-level extractor producing $388{,}105$
declaration nodes and $11.3$M typed edges across 25 Lean projects. We bridge the two
graphs by embedding generated natural-language slogans into a shared semantic space,
linking related statements across papers and across the informal/formal divide; an
LLM judge affirms $47{,}952$ such matches above a $0.8$ cosine floor, with the judge-acceptance rate rising from $48\%$ across the floor to $87\%$ in the $\geq\!0.9$ tier. On formal
concept retrieval, our name-and-signature representation with graph expansion comes within 0.5pp of LeanSearch~v2's reranked Recall@10 ($0.775$ vs.\ $0.780$) without an LM reranker. We release the dataset, extractors, HTTP API, and MCP interface as infrastructure for mathematical search, attribution, and retrieval-augmented reasoning, available at \href{https://www.theoremsearch.com/}{theoremsearch.com} and \href{https://huggingface.co/datasets/uw-math-ai/theorem-matching}{huggingface.co}.
\end{abstract}

\section{Introduction}
Mathematical knowledge is built upon discrete statements and their dependencies. Beginning with a chosen set of axioms, mathematicians derive new statements by applying inference rules; once established, these statements become premises for subsequent results. Yet in contemporary mathematical research, the vast scale and specialization of the literature make these dependency structures difficult to track. Authors typically cite papers rather than the exact lemmas, definitions, or theorems required for a proof, in part because expert readers are expected to supply the relevant background. This practice avoids the burdens of explicit dependency annotations, but coarsens the structure of mathematical knowledge. As a result, attribution may be obscured or rendered imprecise, and related work may be duplicated when researchers are unaware of equivalent or overlapping results. In a large study of withdrawn arXiv submissions, $2.4\%$ of categorized withdrawals are ones whose authors self-identify as ``not novel'' ($357$ of $14{,}839$ cases), reflecting work later judged to duplicate or lack originality relative to existing results \cite{rao2024withdrarxivlargescaledatasetretraction}.

Proof assistants have emerged as an important tool for making mathematical verification more rigorous and explicit. These systems allow mathematicians to express definitions, theorems, and proofs in a formal language whose correctness can be verified by a trusted kernel. In contrast to informal proofs, where references to prior results and routine intermediate arguments are often left implicit, formal proofs must specify each logical dependency and enough intermediate structure to be type-checked by the kernel. As a result, formalization reveals the fine-grained dependency structure underlying mathematical arguments. 

Lean is one of the most prominent modern proof assistants, based on the calculus of constructions and supported by an active open-source ecosystem. Lean's mathematical library, \verb|mathlib4|, contains over 400,000 formalized definitions, lemmas, theorems, and proofs contributed by the community across mathematical fields such as algebra, topology, analysis, and number theory \cite{lean4export}. Despite this rapid growth, formalized mathematics remains much smaller and less comprehensive than the body of informal mathematical knowledge. Lean’s coverage is strongest in foundational material, standard undergraduate- and graduate-level mathematics, and a growing number of targeted research-level formalization projects. Its limited coverage of all research literature caps the extent to which formal methods can be used to navigate or verify contemporary mathematics at scale.

In this work, we collect and analyze dependency graphs for both formalized and informal mathematics. Our main contributions are: 
\begin{enumerate}
    \item \textbf{Informal dependency graph.} We extract dependency structure from informal mathematical sources, including arXiv and Lean Community, recovering 18.3 million dependencies connecting over 11.7 million statements.
    \item \textbf{Formal dependency graph.} We extract typed declaration-level dependencies from Mathlib4 and several open-source formalization projects.
    \item \textbf{Systematic analyses.} We perform comprehensive studies of both graphs, examining their structural properties and their applicability to mathematical search, navigation, and reasoning.
    \item \textbf{Cross-formality bridge.} We link informal and formal statements in a shared slogan-embedding space, yielding $47{,}952$ LLM-affirmed (informal, formal) matches; the same name-and-signature representation with graph expansion comes within 0.5pp points of LeanSearch~v2's reranked Recall@10 without a reranker.
    \item \textbf{Public dataset.} We release the resulting formal and informal dependency data as a public resource for the mathematical and machine-learning communities.
    \item \textbf{Programmatic access.} We provide an API and MCP interface that allow users, applications, and AI agents to query the dependency data directly.
\end{enumerate}

\section{Related Work}

\subsection{Mathematical Information Retrieval}

Early Math IR focused on formula-aware and keyword search, motivated by the fact that mathematical notation cannot be treated as ordinary text. Benchmarks such as NTCIR-11 Math-2, NTCIR-12 MathIR, and ARQMath formalized retrieval over mathematical corpora containing both formulas and surrounding text, including arXiv, Wikipedia, and Math Stack Exchange \cite{ntcir-11-math-2, ntcir-12-math-ir, mansouri-arqmath-3}. These settings showed that mathematical meaning depends on both symbolic structure and visual layout, motivating representations such as Symbol Layout Trees and Operator Trees with formula-specific indexing methods \cite{zanibbi2015tangentsearchengineimproved}. More recent work explores learned representations: zbMath-BERT studies embeddings for mathematical text \cite{zbmath-bert}, NaturalProofs and TheoremSearch apply neural retrieval to informal statements \cite{naturalproofs, theoremsearch2026}, and LeanSearch retrieves Mathlib declarations from informal queries via sloganized formal statements \cite{leansearch-v1, leansearch-v2}. Our work extends sloganized statement retrieval by incorporating explicit dependency structure among mathematical statements.

\subsection{Bibliometrics}

Bibliometric work uses scholarly signals to measure influence, relatedness, and field structure. Citation graphs model relationships among papers, authors, and journals; citation-count measures such as the h-index and journal impact factor summarize influence directly, while recursive prestige models such as Pinski--Narin influence, Eigenfactor, CiteRank, and \.{Z}yczkowski's weighted impact factors weight citations by graph structure, recency, or linking behavior \cite{journal-impact-factor, h-index-paper, pinski-narin, bergstrom-eigenfactor, walker-citerank, zyczkowski-citation-graph}. In mathematics, graph-based analyses have been applied to Mathlib dependency structure and proof-assistant libraries as heterogeneous networks \cite{li-network-structure-mathlib, bauer-datasets-for-formal}, while TheoremKB and AutoMathKG construct theorem-level graphs for informal sources at smaller scale \cite{mishra-theoremkb, automathkg}. Our work extends these graph-based perspectives to theorem-level dependencies across both informal and formal mathematics.

\subsection{Neural Theorem Proving \& Autoformalization}
\label{sec:ntp-autoform}
Neural theorem proving and autoformalization address complementary stages in converting informal mathematics into machine-checked proofs: autoformalization translates informal statements into formal declarations, while theorem proving generates proofs for those declarations. Recent progress on benchmarks such as miniF2F and PutnamBench has increased the importance of retrieval and tooling for agentic proving systems \cite{minif2f-v1, minif2f-v2, tsoukalas2024putnambenchevaluatingneuraltheoremprovers, requena2026minimalagentautomatedtheorem, liu2026numinaleanagentopengeneralagentic}. Our work supports this direction by aligning informal and formal statements with their dependency context.


\subsubsection{Lean Graph}
\label{sec:leangraph}
Dependency extraction for formal libraries is well studied \cite{alama-formal-deps}, and several Lean tools build declaration-level graphs. Jixia, the extractor behind LeanSearch~v2 \cite{jixia}, emits a raw post-elaboration constant-reference graph; Lean~Atlas \cite{yanahama2026leanatlas} splits type- and value-level edges and adds confidence scores, \texttt{sorry}-status, reachability, and visualization; and \texttt{leanblueprint} \cite{massot2020leanblueprint} provides author-annotated \texttt{\textbackslash lean\{\}} links between informal LaTeX statements and their formal Lean declarations. LeanGraph is complementary: it classifies dependencies through Lean's kernel APIs, filters kernel artifacts with a documentation-aligned node set, and refines prior edge taxonomies into \texttt{sig}/\texttt{extends}/\texttt{field} and \texttt{proof}/\texttt{def} edges at multi-library scale.

\section{Informal Graph}

We parse over 11.7 million theorem-like statements from mathematics arXiv papers and recover 18.3 million directed dependency edges within and across papers.

\paragraph{Corpus and statement extraction.}
Our corpus is the arXiv Kaggle metadata snapshot filtered to mathematics-tagged papers. For each paper, we store metadata, references resolved through Semantic Scholar when possible, and the paper's \LaTeX{} source. A regex-based parser identifies theorem-like environments and records each statement's type, reference number, \verb|\label| key, body, proof, and local context. We discard malformed or implausible statements, including empty bodies, unbalanced math delimiters, very short statements, and statements ending mid-clause.

\begin{table}[ht]
  \centering
  \setlength{\tabcolsep}{4pt}
  \small
  \begin{tabular}{lrrr}
    \toprule
    \textbf{Extractor} & \textbf{Edges} & \textbf{Within-paper} & \textbf{External} \\
    \midrule
    Deterministic & 5.23M  & 3.31M  & 1.92M\ (79K) \\
    Heuristic     & 6.47M  & 4.77M  & 1.70M\ (53K) \\
    Notation      & 7.88M  & 7.88M  & -- \\
    \midrule
    Any           & 18.32M & 14.86M & 3.46M\ (130K) \\
    \bottomrule
  \end{tabular}
  \caption{Breakdown of informal dependency edges by extractor. The Within-paper and External columns are not a partition of all edges and need not sum to the Edges total. Per-extractor rows may overlap; the ``Any'' row deduplicates.
  }
  \label{tab:informal-edge-breakdown}
\end{table}

\paragraph{Dependency extraction.}
Each edge is tagged with the extractor or extractors that proposed it (Table~\ref{tab:informal-edge-breakdown}). The \emph{deterministic} extractor resolves \verb|\ref|-like commands within a paper and \verb|\cite| keys against the paper's reference list, using arXiv IDs or title matching; citations that name a target theorem are further resolved to a specific statement. The \emph{heuristic} extractor adds edges from nearby references, backward discourse cues, and prose references such as ``Theorem 3.2,'' with all heuristic edges constrained to point backward in document order. The \emph{notation} extractor uses Qwen3-235B-A22B-Instruct-2507 to identify notation defined and used by each statement, then links each use to the closest prior compatible definition.

\paragraph{Validation.}
We evaluate the extractor pipeline on 500 sampled papers using an independent LLM judge, Kimi K2.5. The judge first proposes within-paper dependencies from the full statement list, then verifies extractor-proposed edges it missed. Results are shown in Table~\ref{tab:informal-dep-judge}; examples of rejected and missed edges appear in Appendix~\ref{app:informal-dep-fails}. Each released edge retains its extractor label, allowing users to trade coverage for precision; deterministic edges, for example, achieve 98.8\% judge-verified precision.

\begin{table}[ht]
  \centering
  \setlength{\tabcolsep}{3pt}
  \small
  \begin{tabular}{lrrr}
    \toprule
    \textbf{Extractor} & \textbf{Edges} & \textbf{Judge-Verified} & \textbf{Judge-Rejected} \\
    \midrule
    Deterministic & 5{,}051  & 4{,}989\ (98.8\%) &      62\ \,(1.2\%) \\
    Heuristic     & 4{,}616  & 3{,}535\ (76.6\%) & 1{,}081\ (23.4\%) \\
    Notation      & 6{,}241  & 2{,}665\ (42.7\%) & 3{,}576\ (57.3\%) \\
    \midrule
    Any           & 14{,}481 & 9{,}855\ (68.1\%) & 4{,}626\ (31.9\%) \\
    \midrule
    \multicolumn{3}{l}{Judge-only edges (no extractor recovered)} & 6{,}372 \\
    \multicolumn{3}{l}{$F_1 \;=\; 2\,\mathrm{TP} / (2\,\mathrm{TP} + \mathrm{FP} + \mathrm{FN})$} & $0.642$ \\
    \bottomrule
  \end{tabular}
  \caption{Agreement between extracted informal dependency edges and an independent LLM judge on 500 arXiv papers. Assuming the judge to be ground truth, judge-verified edges are true positives, judge-rejected extractor edges are false positives, and judge-only edges are false negatives. Per-extractor rows may overlap; the ``Any'' row deduplicates.}
  \label{tab:informal-dep-judge}
\end{table}

\section{Formal Graph}

LeanGraph is a Lean 4 elaborator-level extractor that builds typed declaration-dependency graphs from compiled Lean projects. Each human-facing declaration becomes a node, and dependencies are assigned one of six semantic edge types. The Mathlib v4.27--v4.29 graph contains 351K declaration nodes and 9.3M within-library edges; across all 25 Lean projects, LeanGraph extracts 388,105 nodes and 11,335,708 edges in total (including cross-library).

\paragraph{Scope and granularity.}
LeanGraph runs inside Lean 4 over the kernel \texttt{Environment} API, extracting dependencies from elaborated, type-checked declarations rather than source text. This gives access to post-elaboration constants and proof terms that source-level parsers, such as \texttt{lean4export}'s Python pipeline \cite{lean4export}, cannot reliably recover.

\paragraph{Edge types.}

LeanGraph emits six edge categories. \texttt{extends} captures structure or class inheritance; \texttt{field} captures dependencies in structure field types; \texttt{sig} captures constants appearing in declaration signatures; \texttt{proof} captures constants used in theorem proof terms; \texttt{def} captures constants used in non-\texttt{Prop} definition bodies; and \texttt{docref} captures valid backtick references in docstrings. This taxonomy separates structural, type-level, value-level, and documentation-level dependencies, enabling downstream systems to rank or filter edges by role.

\paragraph{Node inclusion.}
A declaration is included when it satisfies \texttt{shouldIncludeConstant}, a structural predicate based on Lean's classification APIs and doc-gen4's renderability criteria \cite{docgen4}. This removes kernel-generated artifacts such as auxiliary recursors, \texttt{noConfusion} lemmas, matcher declarations, projections, and parent accessors, while retaining user-facing declarations. Anonymous instances and tactic objects are kept but tagged with metadata, allowing downstream consumers to filter them if needed.

\section{Universal Graph}
\label{sec:universal-graph}

The informal and formal graphs differ sharply in density and representation: the informal graph is sparse (edge-to-statement ratio $\sim$1.56), the formal graph is dense ($\sim$29.2), and no edges initially connect them. To link equivalent or closely related statements across papers and across the informal/formal divide, we map every statement into a shared semantic space using slogans, embeddings, and nearest-neighbor lookup.

\paragraph{Slogans.} Following prior work on statement-level retrieval \cite{leansearch-v1, theoremsearch2026}, for each statement, we generate a \emph{slogan}: a concise, standalone natural-language summary produced by Qwen3-235B-A22B-Instruct-2507 \cite{qwen3technicalreport}. Informal slogans are generated with an escalating prompt chain that adds context only when the previous attempt is self-flagged as insufficient. 

The stages add, in order, the statement body, proof and local context, neighboring statements and outgoing dependencies, and finally a forced best-guess slogan. This achieves full coverage over 11.75M statements, with 70.3\% resolved at the minimal stage and 98.1\% before the final fallback (Table~\ref{tab:slogan-chain}). For formal declarations, we rank outgoing dependencies by semantic role and pack the highest-priority mathematically informative context into a fixed-size prompt, filtering ubiquitous declarations and Lean/kernel plumbing. Exact ranking and filtering rules are given in the released code.

\begin{table}[h]
  \centering
  \setlength{\tabcolsep}{4pt}
  \small
  \begin{tabular}{lrrr}
    \toprule
    \textbf{Stage} & \textbf{Statements} & \textbf{Sufficient} & \textbf{Coverage} \\
    \midrule
    \texttt{minimal}       & 11.75M & 8.26M\ (70.3\%) & 70.3\% \\
    \texttt{standard}      & 3.49M  & 2.87M\ (82.1\%) & 94.7\% \\
    \texttt{comprehensive} & 624K   & 406K\ (65.1\%)  & 98.1\% \\
    \texttt{final}         & 218K   & 218K\ (100\%)   & 100\% \\
    \bottomrule
  \end{tabular}
  \caption{Slogan rescue rates along the escalating prompt chain.}
  \label{tab:slogan-chain}
\end{table}

\label{sec:embeddings-semantic-representation}
\paragraph{Embeddings.} Each slogan is embedded into $\mathbb{R}^{4096}$ using Qwen3-Embedding-8B \cite{zhang2025qwen3embedding} with a retrieval instruction, then $\ell_2$-normalized so inner products equal cosine similarity. Vectors are stored in \texttt{pgvector} with an HNSW index \cite{malkov2018efficientrobustapproximatenearest}; a binary-quantized projection is used for fast candidate generation.

\paragraph{Semantic representation.} For each target statement, we can retrieve semantically similar statements using the embedding index. Retrieval first uses binary HNSW to produce an approximate shortlist, then reranks candidates by full-precision cosine similarity and keeps matches above a     fixed threshold. Figure \ref{fig:match-types} details the possibilities: formal (or informal) to itself can surface duplicates while formal to informal finds semantic restatement edges (matches).

\section{Bridging The Corpora}
\label{sec:bridging}

\newcommand{\nBlueprintPairs}{1{,}595\xspace}
\newcommand{\nBlueprintFormal}{1{,}577\xspace}
\newcommand{\nBlueprintInformal}{1{,}308\xspace}
\newcommand{\nSwept}{385{,}657\xspace}
\newcommand{\pctCandidate}{55.3\%\xspace}
\newcommand{\nJudged}{100{,}799\xspace}
\newcommand{\matchTotal}{47{,}952\xspace}
\newcommand{\matchRateAll}{48\%\xspace}
\newcommand{\rateNinety}{87\%\xspace}
\newcommand{\nNinety}{6{,}353\xspace}
\newcommand{\yieldSeventy}{23\%\xspace}
\newcommand{\yieldSixty}{2\%\xspace}
\newcommand{\dsMatch}{61{,}234\xspace}
\newcommand{\dsRate}{61\%\xspace}

\paragraph{Motivation.} To bridge the formal and informal graphs into a Universal Graph, we look for \textbf{matches}: (informal, formal) statement pairs that express the same mathematical result under the same assumptions. A Lean blueprint~\citep{massot2020leanblueprint} is a \LaTeX{} document that writes out a development's definitions, statements, and proofs in human-readable form, and links each statement to the Lean declaration that formalizes it through author-written \texttt{\textbackslash lean\{\}} annotations. These links give us a small set of such matches\footnote{\nBlueprintPairs{} (informal, formal) statement pairs drawn from blueprints, across \nBlueprintInformal{} informal statements and \nBlueprintFormal{} formal declarations; one informal statement is often formalized as several Lean lemmas, which is why formals outnumber informals.}. Extending this set and finding such links at scale better connects Lean to the informal mathematics it formalizes, and gives neural theorem provers and auto-formalization systems an accurate scaffold by which new formalizations can come about.

\begin{figure*}[t]\centering
    \includegraphics[width=\textwidth]{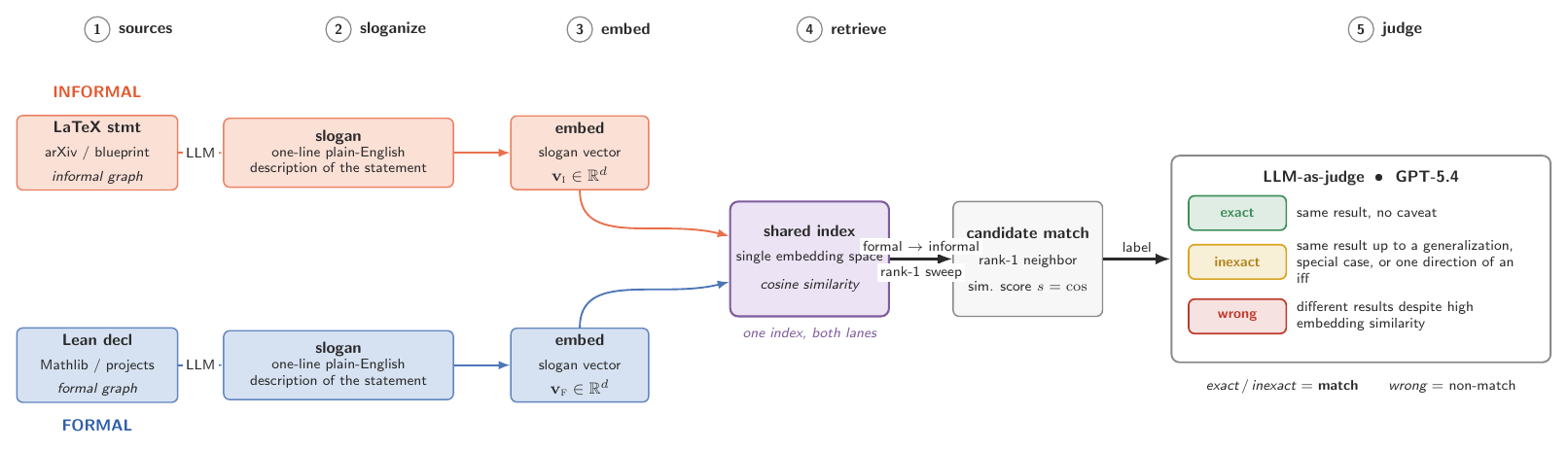}
    \caption{The matching pipeline. Lean declarations and \LaTeX{} statements are each
    sloganized and embedded into one shared index; a formal$\to$informal rank-1 cosine query
    proposes candidate matches, which a single GPT-5.4 judge labels \texttt{exact} /
    \texttt{inexact} / \texttt{wrong} (\texttt{exact}/\texttt{inexact} count as matches).}
    \label{fig:pipeline}
\end{figure*}

\paragraph{Setup.} Our matching pipeline (Figure~\ref{fig:pipeline}) searches from the smaller, lightly-filtered\footnote{We drop constructors (\texttt{ctor}) and compiler-internal auto-generated names, which lack a meaningful natural-language counterpart. This amounts to 2{,}448 nodes ($0.63\%$) and leaves the \nBlueprintPairs{} blueprint pairs intact.} formal graph of \nSwept{} declarations into the informal corpus. For each formal declaration we embed its natural-language slogan and retrieve the single most similar informal slogan by cosine similarity; that nearest neighbor, with its similarity score, is a \textbf{candidate match}.\footnote{Both graphs are embedded into one shared index, so this is a cross-modal nearest-neighbor query, not a per-side search.} At \texttt{ann\_k=50} this finds a candidate for \pctCandidate{} of declarations and yields the similarity distribution in Table~\ref{tab:sim-bins}.

Modulating \texttt{ann\_k} between $500$ and $1000$ over a sample of $200$ otherwise-unmatched declarations, we see that deeper search does recover a candidate for $84\%$ and $88\%$ of them respectively, but only up to a similarity of $0.837$ and $0.859$. Across all $400$ of these deeper probes only a single candidate clears the $0.85$ match bar and none reach $0.90$, so a missing candidate at \texttt{ann\_k=50} almost always reflects the genuine absence of a close informal counterpart rather than a search-horizon limitation.

\definecolor{soft}{HTML}{EAF0F8} 

\begin{table}[t]\centering\small\setlength{\tabcolsep}{6pt}
    \begin{tabular}{@{}l r@{}}\toprule
    Similarity bin & Decls. \\\midrule
    \rowcolor{soft}0.95--1.0  & 678 \\
    \rowcolor{soft}0.90--0.95 & 6{,}636 \\
    \rowcolor{soft}0.85--0.90 & 26{,}807 \\
    \rowcolor{soft}0.80--0.85 & 66{,}744 \\
    0.50\textsuperscript{*}--0.80 & 112{,}481 \\
    no match  & 172{,}311 \\\midrule
    Total      & 385{,}657 \\\bottomrule
    \end{tabular}
    \caption{\textbf{Candidate match} cosine similarity across all 385,657 swept formal declarations. Shaded rows are the tiers evaluated using LLM-as-judge. 0.50* is the minimum retrieved similarity and no match represents declarations with no informal neighbor at \texttt{ann\_k=50}.}
    \label{tab:sim-bins}
\end{table}

\paragraph{Does retrieval find true counterparts?} Embedding formal and informal slogans into one shared space is what makes cross-modal matching possible: a formal statement can be paired with its informal counterpart directly by embedding similarity, with no shared vocabulary or surface overlap required. We validate that this similarity actually recovers true counterparts using blueprint pairs as a source of truth, where each blueprint annotates a Lean declaration with a human-written informal statement. The informal halves sit in our corpus like any other slogan (labeled \texttt{source=blueprint}), with the same embeddings and none treated specially.

We take each blueprint's \emph{formal} slogan as the query and ask which informal slogan is most similar to it in the shared index, then check where the annotated informal partner lands in that ranking. The partner ranks first 43.5\% of the time and within the top ten 69.9\% (Hit@1, Hit@10). Mismatches with the annotated partner are mostly not errors: among formal nodes whose top result is \emph{not} the annotated partner but is itself a high-similarity candidate ($\geq 0.85$), LLM-as-judge rates 65\% (n=187) as matches, i.e.\ valid alternative restatements rather than retrieval failures.

\paragraph{Judging.} To judge matches, a GPT-5.4 judge labels each candidate, given the declaration name, both slogans, the Lean signature, the arXiv title, and the informal \LaTeX. The judge returns its decision as one label with a brief rationale:
\begin{itemize}
\setlength{\itemsep}{1pt}
\item \bx{green!20}{\texttt{exact}}: the same result without caveat; a mathematician could cite one as the other.
\item \bx{yellow!35}{\texttt{inexact}}: the same underlying result but not verbatim (a generalization, special case, or one direction of a bidirectional implication).
\item \bx{red!22}{\texttt{wrong}}: different results despite high embedding similarity (shared vocabulary or sibling concepts).
\end{itemize}
We count \texttt{exact} and \texttt{inexact} as matches. GPT-5.4 is our judge of record because it is stricter than our initial choice, Opus~4.8, which a domain expert found over-generous; since a curated set should favor precision, we adopt the stricter judge (selection and an expert re-grading are in Appendix~\ref{app:calib}). The judge is stable under re-runs: a second independent GPT-5.4 pass over a random 500-candidate sample agrees on the match/non-match verdict for $93.2\%$ of pairs ($\kappa{=}0.86$)\footnote{Cohen's $\kappa$ corrects agreement for chance: $\kappa = (p_o - p_e)/(1 - p_e)$, where $p_o$ is the observed agreement and $p_e$ the agreement expected if the two raters labeled independently at their observed match rates.}, so re-running GPT-5.4 rarely changes a verdict.

\begin{figure*}[t]
  \centering
  \begin{subfigure}{0.49\linewidth}
    \centering
    \includegraphics[width=\linewidth]{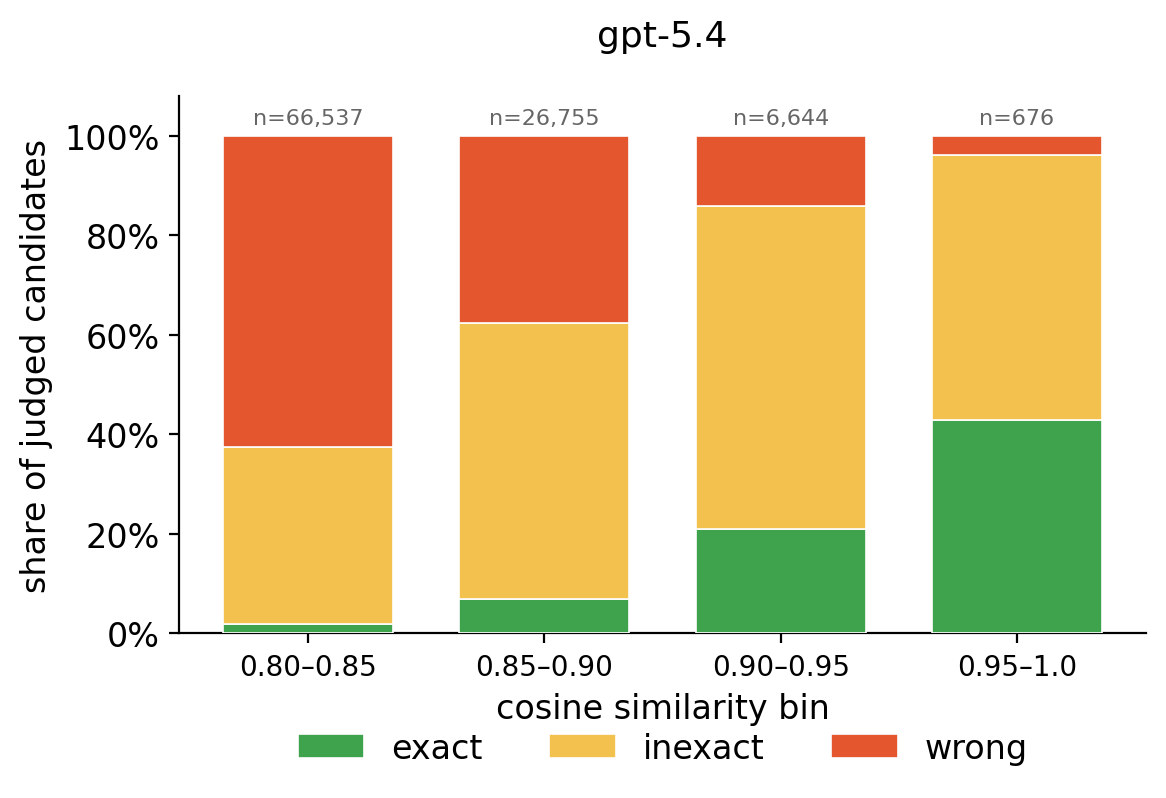}
    \caption{GPT-5.4, our judge of record.}
    \label{fig:comp-g54}
  \end{subfigure}
  \hfill
  \begin{subfigure}{0.49\linewidth}
    \centering
    \includegraphics[width=\linewidth]{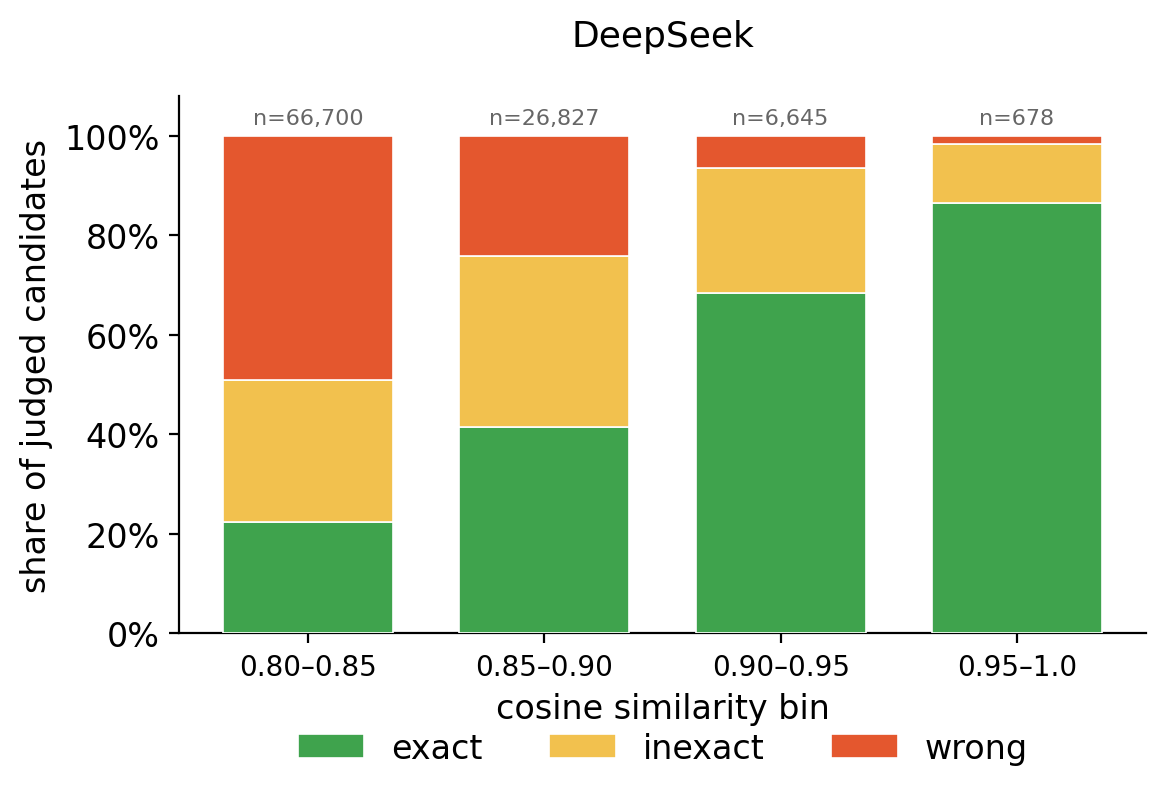}
    \caption{DeepSeek, a more lenient high-recall comparison.}
    \label{fig:comp-ds}
  \end{subfigure}
  \caption{Verdict composition by cosine-similarity bin (candidates at $\geq\!0.8$). Both judges shift from \texttt{wrong} to \texttt{exact} and \texttt{inexact} as similarity rises; DeepSeek is consistently more generous, so we keep GPT-5.4 as the stricter set of record and report DeepSeek as a high-recall comparison.}
  \label{fig:comp}
\end{figure*}

\paragraph{Where to judge.} We judge every candidate at cosine similarity $0.8$ and above. We set the floor here, rather than higher, to make the curated set as valuable as possible: it captures the broad band of \texttt{inexact} matches alongside the \texttt{exact} ones, and because each edge is stored with its deterministic graph context, later work can re-judge these looser links with more information and promote \texttt{inexact} ties to \texttt{exact}. We do not go lower because the \textbf{yield rate} (the fraction the judge affirms as a match) collapses below $0.8$: a pilot of 150 candidates per $0.1$-wide bin confirms only \yieldSeventy{} in $0.7$--$0.8$ and \yieldSixty{} in $0.6$--$0.7$, where judging mostly confirms non-matches. Above the floor the match rate climbs steeply with similarity (Figure~\ref{fig:comp}).

\paragraph{Judged matches.} Of the \nJudged{} candidates above the floor\footnote{The retrieval pool holds 100{,}865 candidates at similarity $\geq 0.8$ (Table~\ref{tab:sim-bins}); 100{,}831 were submitted to GPT-5.4 and 32 returned unparseable output, giving these 100{,}799 valid labels. Of them, 187 are \texttt{unjudgeable}, leaving 100{,}612 with a decisive \texttt{exact}/\texttt{inexact}/\texttt{wrong} verdict.}, GPT-5.4 affirms \matchTotal{} as matches (\matchRateAll). Quality concentrates at the high-similarity end: the $\geq\!0.9$ tier alone gives \nNinety{} matches (\rateNinety{} of its candidates), and the affirmation rate rises steadily with similarity across the band (Figure~\ref{fig:comp}). Running a second, more lenient judge (DeepSeek-V4-Pro) over the same candidates affirms \dsMatch{} (\dsRate); we keep the stricter GPT-5.4 set as our matches and report DeepSeek as a high-recall comparison.

\paragraph{Looking to the future.} Embedding both graphs into one space turns cosine similarity into a signal of ``warmth'' that curates high-quality links across two large and largely disconnected bodies of mathematical writing. These links are hard to surface by hand yet cheap to check once proposed, and they give automated formalization systems concrete context to build on, speaking to the ``dependency issue'' Seed-Prover~1.5 describes, where progress ``typically hinges on synthesizing insights across a multitude of related papers''~\citep{chen2025seedprover15masteringundergraduatelevel}.

\section{Retrieval-Augmented Formalization}
\label{sec:exp3}

We test whether slogan retrieval helps an LLM autoformalize natural-language statements into Lean. We use a statement-only setting: given an informal description, \texttt{claude-sonnet-4-6} generates a single Lean~4 declaration ending in \texttt{:= sorry}, with no proof. The compiler is available during generation, with at most three \texttt{typecheck} calls per target.

To reduce memorization, we evaluate on 24 theorems introduced in Mathlib v4.30 while retrieving only from the v4.29 corpus. Queries are \texttt{qwen3-8b} back-translations of the gold Lean signature with the declaration name removed, so the informal text must identify the relevant mathematical objects without relying on the target name. The retriever uses slogan representations with a query head trained on typed dependency labels; on a held-out Mathlib premise benchmark, this improves R@100 from 0.16 with raw cosine similarity to 0.54. On the 24-target evaluation, top-15 retrieval recall is 0.161.

We compare four conditions: \emph{None}, using only the informal statement; \emph{RAG}, adding the top retrieved premises; \emph{Library}, adding a \texttt{grep} tool over the full v4.29 declaration listing; and \emph{RAG+Library}, allowing both. A \texttt{claude-opus-4-7} judge, checked by hand, labels each output as \emph{strict} if it states the same proposition up to notation, and \emph{evaluated correct} if it is a high-confidence equivalent restatement.

\begin{table}[t]
  \centering
  \scriptsize
  \setlength{\tabcolsep}{3.5pt}
  \begin{tabular}{lccccc}
    \toprule
    \textbf{Cond.} & \textbf{TC} & \textbf{Strict} & \textbf{Eval.} & \textbf{Tok.} & \textbf{Calls} \\
    \midrule
    None          & 22/24 & 4/24 & 5/24 & 52k & 276 \\
    RAG           & 20/24 & 6/24 & \textbf{8/24} & \textbf{14k} & \textbf{68} \\
    Library       & 23/24 & 5/24 & 6/24 & 52k & 275 \\
    RAG+Library   & \textbf{24/24} & \textbf{7/24} & \textbf{8/24} & 37k & 188 \\
    \bottomrule
  \end{tabular}
  \caption{Statement-only autoformalization of 24 Mathlib v4.30 theorems, retrieving over v4.29.
  \emph{Strict} = same proposition; \emph{Eval.} = strict plus high-confidence equivalent
  (\texttt{claude-opus-4-7}-judged, hand-checked). \emph{TC} = typecheck.}
  \label{tab:formalize}
\end{table}

\paragraph{Results.} Retrieval improves evaluated correctness from 5/24 to 8/24 while using fewer output tokens and tool calls than library search (Table~\ref{tab:formalize}). The RAG condition matches RAG+Library on evaluated correctness, but uses 14k output tokens and 68 tool calls, compared with 37k tokens and 188 calls. Typechecking alone is not a reliable success signal: the ungrounded condition typechecks 22/24 outputs but is evaluated correct on only 5/24. About 10/24 targets fail in every condition, suggesting that some informal queries underdetermine the intended Lean statement.


\section{Comparison Against Existing Retrieval}
\label{sec:exp4_lsv2}

We benchmark our retrieval system against LeanSearch-v2 (LSv2) \citep{leansearch-v2}, a Lean-specific retriever. Both systems retrieve flat embedded passages by cosine similarity over the same off-the-shelf embedder (Qwen3-Embedding-8B, 4096-dim, normalized; no fine-tuning), and differ only in what those passages contain. LSv2 embeds a structured passage (kind, type signature, and informalized description, plus the value body for definitions, classes, and instances) and keeps the fully-qualified name as metadata alongside it; our baseline embeds the slogan alone.

Because the embedder is held fixed, differences trace to representation rather than embedding-model strength. Two things still differ: the structured fields each side embeds, and the informal text itself (our one-line slogan versus LSv2's dependency-grounded description). Our interventions target the first and do not isolate the contribution of informal-text quality. We evaluate this baseline on MathlibQR, a benchmark from the LSv2 team, and ask whether targeted interventions close the gap: once our passages carry the same name and signature information as LSv2's, our pipeline matches LSv2's full system on recall without reranking.

\paragraph{Benchmark.}
MathlibQR is an informal-to-formal retrieval benchmark that pairs 200 Mathlib declarations with up to 6 query styles each: plain English, \LaTeX{}, Lean-flavored text, slogan, nickname, and a special case. Each is scored by two ranking metrics\footnote{Recall@10 is the fraction of queries whose target declaration appears anywhere in the top 10 retrieved results, ranked by similarity. nDCG@10 is a rank-weighted score over the top 10 that rewards ranking the target declaration higher, normalized so a perfect ranking scores 1.}, Recall@10 and nDCG@10, and we also report the shallower nDCG cutoffs\footnote{We report nDCG@\{1,5,10\} and Recall@10, the cutoffs relevant to interactive search. LSv2 additionally reports Recall@\{50,100\}, which target deep recall for downstream premise feeding rather than top-of-list quality, and an LLM-judge ranking we did not replicate.}.  Because compared systems are built on continuously-evolving Mathlib snapshots, a system can miss a query simply because the target declaration is absent, rather than retrieval failing.

The fair-810 subset described in LSv2 controls for this: its 810 query rows cover 171 of the 200 available target declarations, and each of those targets is guaranteed to appear in every compared system's corpus. To tie our system into this benchmark, we run our own retrieval pipeline over our snapshot of the corresponding Mathlib versions ($\text{v}4.27 \cup \text{v}4.28$), unioned to maximize coverage of the fair-810 targets) and map LSv2's target declarations onto ours; all 171 are present in our index.

\begin{figure}[h!]
    \centering
    \includegraphics[width=0.25\textwidth]{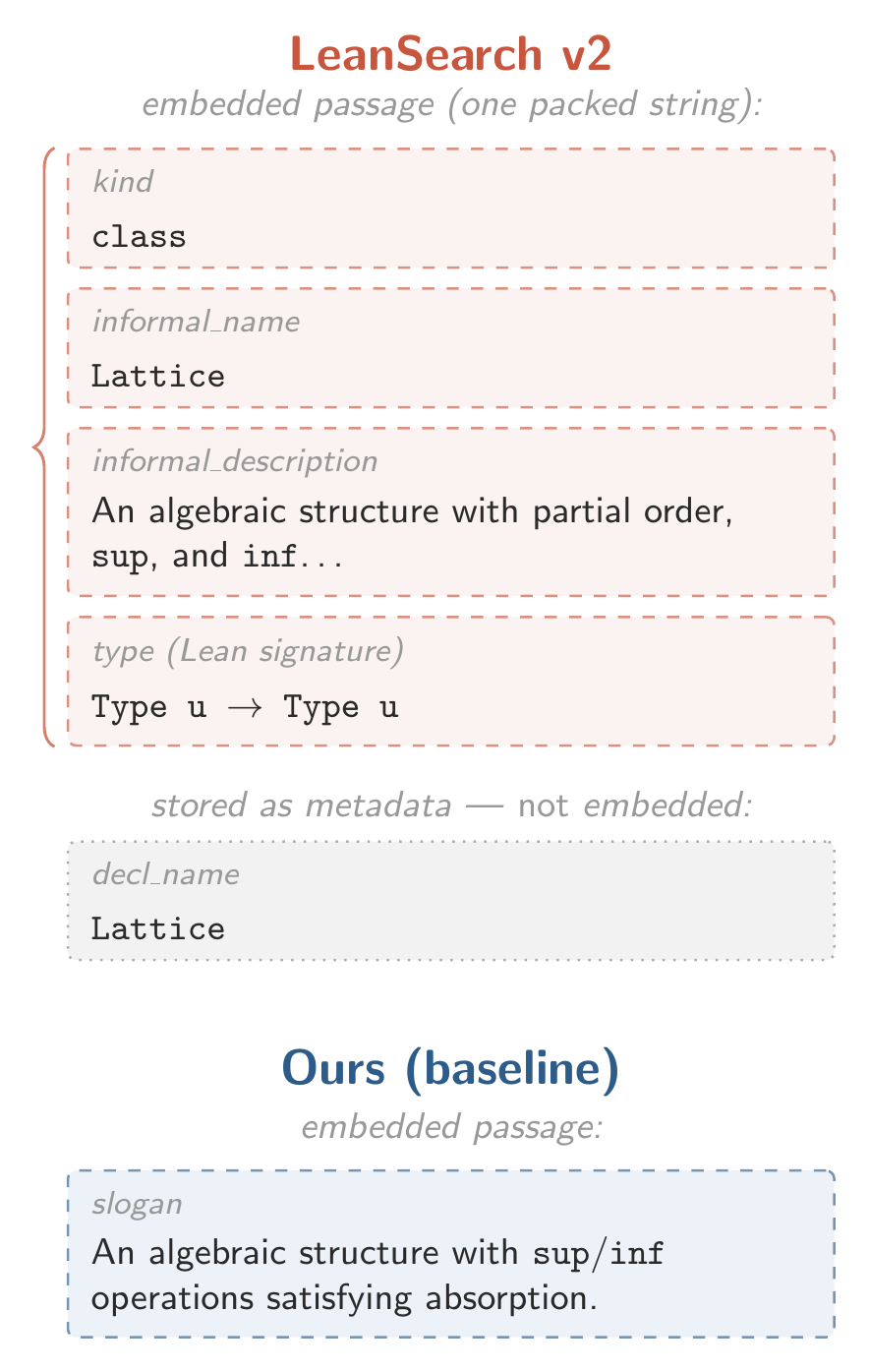}
    \caption{What each system embeds for a declaration (example: the \texttt{Lattice} class). LeanSearch v2 packs declaration kind, informal name and description, and type signature into a single embedded passage, plus the value body for definitions, classes, and instances. The fully-qualified name is kept only as metadata, outside the embedded text~\cite{leansearch-v2}. Our baseline embeds a single natural-language slogan, with neither the signature nor a lexical handle on the name. The name-and-signature representation and name-matching index we introduce below add that channel.}
    \label{fig:lsv2-repr-gap}
\end{figure}

\paragraph{The representation gap.}
Our baseline embeds only the slogan and scores 0.586 Recall@10 / 0.380 nDCG@10, 7.1pp and 11.4pp behind LSv2's retriever-only baseline (Table~\ref{tab:lsv2_ablation}). LSv2's embedded passage carries the type signature and an informalized description, whereas our slogan carries only a one-line gloss (Figure~\ref{fig:lsv2-repr-gap}); for declarations whose identity is essentially their name, our baseline has little to match against. This shows up in a few ways. 

First, some declarations like \texttt{Lattice} or \texttt{Functor} \emph{are} their name, meaning that structures or classes with little informal elaboration are poorly retrieved. In the baseline, these are our two weakest kinds with nDCG@10 0.222 and 0.253 (respectively), against 0.477 for theorems - whose descriptions a slogan captures well. Second, some queries are typed as bare name or in Lean syntax rather than plain English, and the baseline handles these poorly. When both line up, a Lean-style query for a named structure against our baseline finds the right declaration in the top ten only 12.9\% of the time.

\paragraph{Closing the gap.}
We group the interventions by what each improves, the ablation in Table~\ref{tab:lsv2_ablation} adds them cumulatively across Configurations A-E.

\begin{figure}[h!]
    \centering
    \includegraphics[width=\linewidth]{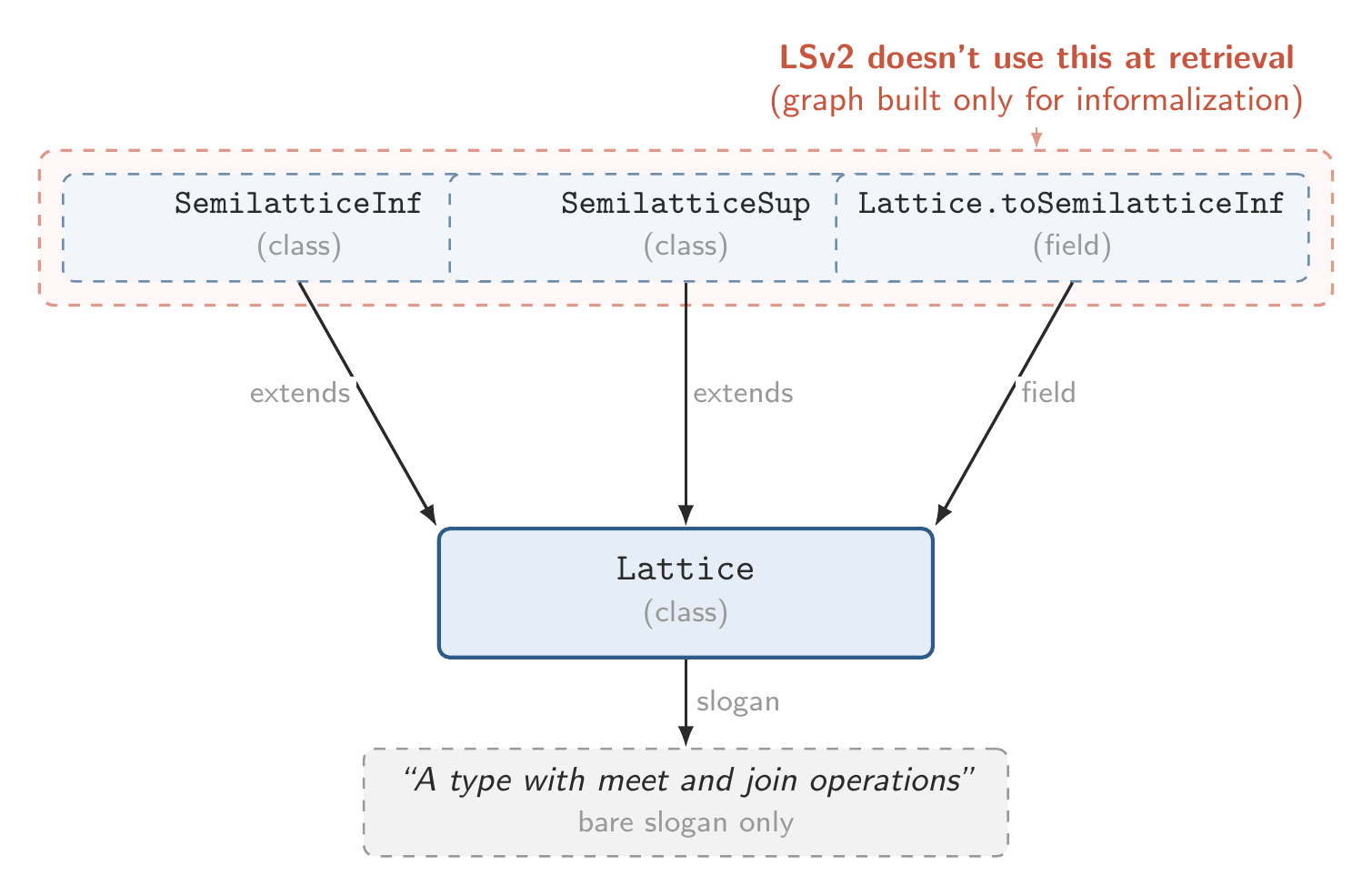}
    \caption{Graph expansion on \texttt{Lattice} (Config~E). After retrieval we follow the dependency graph one hop up and add each candidate's \texttt{formal\_dependency} parents. Their names (\texttt{SemilatticeInf}, \texttt{SemilatticeSup}, \texttt{Lattice.toSemilatticeInf}) match name-like queries the bare slogan misses. LSv2 builds the same graph but uses it only for informalization, not at retrieval.}
  \label{fig:graph-expansion}
\end{figure}

\begin{itemize}
    \item \textbf{Putting the name where queries can reach it} (\emph{Name/sig}; the name-matching index is part of \emph{Search}). We add a name-and-signature representation: a second embedded passage encoding the declaration's name, type signature, and dependencies. This moves the fully-qualified name, which LSv2 keeps only as metadata, into the embedded text itself (Figure~\ref{fig:lsv2-repr-gap}), giving name-like queries an explicit target. We also add a name-matching index over declaration names for the bare-name queries the embedding may miss. Together these directly fix the two error classes from the representation gap.
    \item \textbf{Translating the query into our wording} (\emph{Search}). For queries written tersely or in Lean syntax, which are unlike our slogans, we have a model draft a hypothetical informal statement answering it and search with that alongside the original query. This fuses the two rankings (HyDE), and phrases the search closer to our slogans.
    \item \textbf{Resolving same-name collisions} (\emph{Graph}). After retrieving, we follow the dependency graph one hop up and add each result's parent declarations (Figure~\ref{fig:graph-expansion}), which surfaces the general concept when a more specific declaration of the same name has outranked it. LSv2 builds the same graph and uses it to inform corpus construction, but does not consult it at retrieval time.
    \item \textbf{Keeping good candidates in the running} (\emph{Search}). We widen the binary-HNSW shortlist (\texttt{ann\_k}) so correct declarations survive into the full-precision cosine pass rather than being pruned early.
\end{itemize}

\begin{table*}[t]
  \centering \small \setlength{\tabcolsep}{5pt}
  \caption{MathlibQR fair-810. Component columns show what each of our
  configurations uses: the informal slogan, the name-and-signature representation,
  the search-time techniques (name-matching index, query rewriting, wider
  shortlist), and graph expansion. E is recall-optimized, reaching LSv2's reranked
  recall to within 0.5pp (0.775 vs.\ 0.780) without a reranker; F is ranking-optimized.}
  \label{tab:lsv2_ablation}
  \begin{tabular}{l cccc rrrr}
    \toprule
    & \multicolumn{4}{c}{\textbf{Components}} & \multicolumn{4}{c}{\textbf{Metrics}} \\
    \cmidrule(lr){2-5}\cmidrule(lr){6-9}
    \textbf{Configuration} & Slogan & Name/sig & Search & Graph & nDCG@1 & nDCG@5 & nDCG@10 & R@10 \\
    \midrule
    Ours (A) Baseline                    & \checkmark &            &            &            & 0.201 & 0.345 & 0.380 & 0.586 \\
    Ours (B)                             & \checkmark &            & \checkmark &            & 0.262 & 0.441 & 0.467 & 0.680 \\
    Ours (C)                             & \checkmark &            & \checkmark & \checkmark & 0.319 & 0.461 & 0.504 & 0.716 \\
    Ours (D)                             & \checkmark & \checkmark & \checkmark &            & 0.343 & 0.516 & 0.550 & 0.767 \\
    \textbf{Ours (E)} \emph{recall-optimized}  & \checkmark & \checkmark & \checkmark & \checkmark & 0.349 & 0.504 & 0.548 & 0.775 \\
    \midrule
    \textbf{Ours (F)} \emph{ranking-optimized} &            & \checkmark & \checkmark &            & 0.391 & 0.534 & 0.558 & 0.733 \\
    \midrule
    \multicolumn{5}{l}{LSv2 Retriever}            & 0.340 & 0.472 & 0.494 & 0.657 \\
    \multicolumn{5}{l}{LSv2 Retriever + reranker} & \textbf{0.470} & \textbf{0.601} & \textbf{0.623} & \textbf{0.780} \\
    \bottomrule
  \end{tabular}
\end{table*}

\paragraph{Results}
The full ablation is in Table~\ref{tab:lsv2_ablation}; two configurations are worth highlighting. \emph{Ranking-optimized} Configuration~F replaces the slogan with the name-and-signature representation, keeping the search techniques (name-matching index, query rewriting, wider shortlist) but no graph. Without a reranker, it beats LSv2's retriever-only baseline on every metric we report, reaching \textbf{0.558} nDCG@10 against their 0.494 and \textbf{0.733} Recall@10 against their 0.657. Since B and F share the same search techniques and differ only in what they embed, this swap isolates the representation, rather than the search techniques, as the driver of ranking quality.

Combining the slogan with the name-and-signature representation (Configuration~D) and then adding graph expansion (Configuration~E) trades ranking for recall: Recall@10 climbs from 0.733 (F) to 0.767 (D) to \textbf{0.775} (E), while nDCG@10 slips from 0.558 (F) to 0.550 (D) to 0.548 (E). The recall-optimized configuration, E, comes within 0.5pp of LSv2's reranked recall (0.775 vs.\ 0.780) without a reranker;\footnote{Bootstrap 95\% CIs (5000 resamples): Configuration~E Recall@10 0.775~[0.746, 0.802], so LSv2's 0.780 lies within the interval (a failure to reject, not an equivalence test); gains over baseline \mbox{+18.9pp} Recall@10~[+15.1, +22.7] and \mbox{+16.8pp} nDCG@10~[+13.3, +20.2].} neither E nor F matches the reranked nDCG of 0.623, where the reranker's precise ordering leads at every cutoff.

\begin{figure*}
    \centering
    \includegraphics[width=\linewidth]{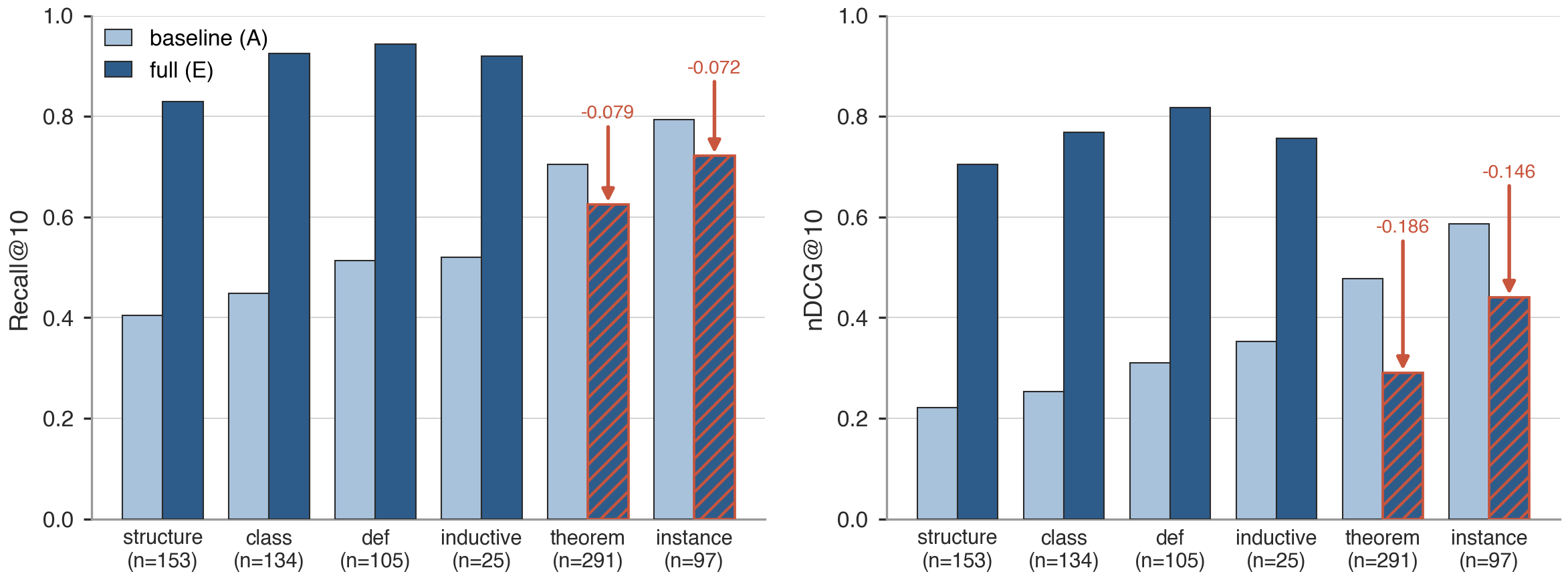}
    \caption{MathlibQR fair-810, Recall@10 by declaration kind, comparing the baseline (Config~A) with the full lever stack (Config~E). The interventions concentrate their gains on declarations identified mainly by name (structures, classes, definitions, and inductives), while \texttt{theorem} and \texttt{instance}, which usually already carry descriptive slogans, decline slightly. Configurations are defined in Table~\ref{tab:lsv2_ablation}.}
    \label{fig:lsv2-perkind}
\end{figure*}

Per declaration type (Figure~\ref{fig:lsv2-perkind}), the levers primarily improve the categories for which slogans were least effective. Structures, classes, definitions, and inductives all benefit, whereas \texttt{theorem} and \texttt{instance}, which typically already have descriptive slogans, decline slightly. Thus, the interventions incur a small cost on well-described declarations in exchange for substantial gains on declarations identified mainly by name. 

Across the same 810 queries the interventions add \textbf{+18.9pp} Recall@10 and \textbf{+16.8pp} nDCG@10 over the baseline, while graph expansion contributes least once the representation is present (E vs.\ D: +0.8pp Recall, $-$0.2pp nDCG).

\paragraph{Transfer to MathlibMPR.}
The same configuration does not carry over to chained-premise retrieval. On MathlibMPR, applying the QR-tuned configuration lowers our group Recall@10 from 0.224 to 0.165 against our untuned baseline, so the interventions that help concept retrieval are counterproductive here. The task asks which lemmas a proof invokes, and those can share no vocabulary with the theorem, whereas every technique above sharpens matching on the meaning of a single declaration. We treat this as a scope boundary, not a comparison, and leave chained-premise retrieval to future work.

\section{Conclusion}
We present TheoremGraph, a statement-level dependency graph linking $11.7$M informal arXiv statements ($18.3$M edges) with the Lean ecosystem ($388{,}105$ declarations across 25 projects, $11.3$M typed edges) through a shared slogan-embedding space. Embedding both corpora into one index turns cross-formality matching into a nearest-neighbor query: an LLM judge affirms $47{,}952$ matches above a $0.8$ cosine floor,  affirmed in $87\%$ of its $\geq\!0.9$-similarity candidates. The same representation improves downstream tasks over lexical and dense-only baselines: it raises retrieval-augmented autoformalization from $5/24$ to $8/24$ evaluated-correct targets, and on MathlibQR fair-810 it comes within 0.5pp of LeanSearch~v2's reranked Recall@10 ($0.775$ vs.\ $0.780$) without an LM reranker. We release the dataset, extractors, HTTP API, and MCP interface as infrastructure for mathematical search and retrieval-augmented reasoning.

\section{Limitations}
TheoremGraph is built from arXiv \LaTeX{} source, a large structured corpus that does not cover all mathematical writing. Informal dependency extraction is necessarily approximate, so the released graph preserves per-extractor labels, letting users pick their own precision--coverage tradeoff.

The cross-formality bridge relies on LLM-generated slogans and Qwen3-Embedding-8B embeddings, and sloganization is lossy. The matching judge works from slogans with limited context, and our expert calibration covers only ten pairs, so not every judge-affirmed match is ground truth. Giving judges more context, such as the source paper, is a natural way to improve cross-corpus linking, alongside evaluating alternative slogan generators, embedding models, rerankers, and a broader judge panel. Because matching takes only the single nearest candidate over an arXiv-only corpus, well-known results go unmatched when their counterpart is absent.

Our downstream evaluations focus on settings with gold labels: blueprint pairs, $24$ autoformalization targets, and the MathlibQR fair-810 subset. The same configuration does not transfer to chained-premise retrieval: applying the QR-tuned settings lowers MathlibMPR group Recall@10 from $0.224$ to $0.165$. Concept retrieval and chained-premise retrieval appear to need different signals, which we leave to future work.

\paragraph{Data availability.} The TheoremGraph dataset, the LeanGraph and informal extractors, and the HTTP API and MCP interface are available at \href{https://www.theoremsearch.com}{theoremsearch.com}. The judged formal--informal matches are released at \href{https://huggingface.co/datasets/uw-math-ai/theorem-matching}{huggingface.co/datasets/uw-math-ai/theorem-matching}. 

Because arXiv's default license grants no third-party redistribution right, the public release is restricted to statements carrying an open, redistributable license: a subset of $23{,}399$ judged candidates, gated for non-commercial research. We license our own contributions (the slogans, judgments, and metadata) under CC-BY-NC-SA-4.0, applied to the compilation as a whole rather than to the underlying statements; each record additionally carries its statement's original license in a \texttt{source\_license} field, which governs reuse of that statement's content.

\bibliography{custom}
\appendix
\onecolumn
\newpage
\section{Example Judge-Rejected and Missed Informal Dependencies}
\label{app:informal-dep-fails}

For each extractor and for the judge, we present three samples drawn from the 500-paper judge subset. The first three tables show edges the parser found but the LLM judge rejected; the fourth shows edges only the judge identified. Statement bodies are reproduced from the parsed \LaTeX\ source.

\begingroup
\small\setlength{\tabcolsep}{6pt}
\begin{longtable}{@{}p{0.48\linewidth}p{0.48\linewidth}@{}}
\caption{Deterministic edges the judge rejected. The first and last are explicit author-declared restatements that the judge rejected. The second makes an explicit reference to a lemma in the proof.}
\label{tab:dep-det} \\
\toprule
\textbf{Source statement} & \textbf{Dependency statement} \\
\midrule
\endfirsthead
\caption[]{Deterministic edges the judge rejected (continued)} \\
\toprule
\textbf{Source statement} & \textbf{Dependency statement} \\
\midrule
\endhead
\bottomrule
\endfoot

\multicolumn{2}{@{}>{\centering\arraybackslash}p{\dimexpr0.96\textwidth+1.5em\relax}@{}}{%
  \itshape From ``Object-unital groupoid graded rings, crossed products and separability''
  (arXiv:2001.05164)%
} \\[2pt]
\textbf{Proposition 48.} \emph{(Theorem 3)} Suppose that $L/K$ is a separable (not necessarily finite) field extension. Then the object crossed product $R = (L/K,\beta)$ is separable over $R_0$ if and only if $\mathrm{Aut}_K(L)$ is finite. In particular, if $L/K$ is Galois, then $R$ is separable over $R_0 = L$ if and only if $L/K$ is finite. & \textbf{Theorem 3.} Suppose that $L/K$ is a separable (not necessarily finite) field extension. Then the object crossed product $R = (L/K,\beta)$ is separable over $R_0$ if and only if $\mathrm{Aut}_K(L)$ is finite. In particular, if $L/K$ is Galois, then $R$ is separable over $R_0 = L$ if and only if $L/K$ is finite. \\
\midrule
\addlinespace
\multicolumn{2}{@{}>{\centering\arraybackslash}p{\linewidth}@{}}{%
  \itshape From ``Koszul duality and equivalences of categories''
  (arXiv:math/0012264)%
} \\[2pt]
\textbf{Corollary 5.9.} Let $P$ be any complex in $\mathrm{KomF}(U)$. (a) $H^p G(P) = H^p(k \otimes_U P)$. (b) $P$ is in $N_{K(U)}$ iff $P$ and $k \otimes_U P$ are acyclic.
& \textbf{Lemma 5.4.} Let $I \in \mathrm{KomCoF}(A^{!},d)$ be (forgetting the differential) $I = \prod_{p > 0} \mathrm{Hom}_k(A^{!}, N^p(-p))$. Then $H^p F(I) = 0$ for $p \leq 0$. \\
\midrule
\addlinespace
\multicolumn{2}{@{}>{\centering\arraybackslash}p{\dimexpr0.96\textwidth+1.5em\relax}@{}}{%
  \itshape From ``Cluster Algebras, Invariant Theory, and Kronecker Coefficients II''
  (arXiv:1711.00163)%
} \\[2pt]
\textbf{Theorem 2.} \emph{(Theorem 2.0.1.22)} There is a rigid potential $\widetilde{W}_l^m$ on $\widetilde{\Diamond}_l^m$ such that $(\widetilde{\Diamond}_l^m, \widetilde{W}_l^m)$ is a polyhedral cluster model. & \textbf{Theorem 2.0.1.22.} The rigid potential IQP $(\overline{\Diamond}_l^m, \overline{W}_l^m)$ is a polyhedral cluster model and its Jacobian algebra is finite-dimensional. \\
\end{longtable}
\endgroup

\begingroup
\small\setlength{\tabcolsep}{6pt}
\begin{longtable}{@{}p{0.48\linewidth}p{0.48\linewidth}@{}}
\caption{Heuristic edges the judge rejected. These illustrate a common heuristic failure: sibling theorems in adjacent positions, where proximity-weighted scoring fires but the two are parallel results rather than a dependency.}
\label{tab:dep-heur} \\
\toprule
\textbf{Source statement} & \textbf{Dependency statement} \\
\midrule
\endfirsthead
\caption[]{Heuristic edges the judge rejected (continued)} \\
\toprule
\textbf{Source statement} & \textbf{Dependency statement} \\
\midrule
\endhead
\bottomrule
\endfoot

\multicolumn{2}{@{}>{\centering\arraybackslash}p{\dimexpr0.96\textwidth+1.5em\relax}@{}}{%
  \itshape From ``On the spectral $\nu$-continuity''
  (arXiv:2009.08977)%
} \\[2pt]
\textbf{Theorem 4.2.} Let $T \in \Phi(H)$ be such that $T$ satisfies Browder's theorem. If $0 \in \mathrm{acc}\,\sigma_w(T)$ and $(T_n)$ is a sequence of essentially $G_1$ operators such that $T_n \xrightarrow{\nu} T$, then $\sigma(T_n) \to \sigma(T)$. & \textbf{Theorem 4.1.} Let $T \in \Phi(H)$ be such that $0 \notin \sigma_w(T)$ or $0 \in \mathrm{acc}\,\sigma_w(T)$. If $(T_n)$ is a sequence of essentially $G_1$ operators such that $T_n \xrightarrow{\nu} T$, then $\sigma_w(T_n) \to \sigma_w(T)$. \\
\addlinespace
\multicolumn{2}{@{}>{\centering\arraybackslash}p{\dimexpr0.96\textwidth+1.5em\relax}@{}}{%
  \itshape From ``Polar Codes for the m-User MAC''
  (arXiv:1002.0777)%
} \\[2pt]
\textbf{Lemma 7.} $U_{2,4}$ cannot be the uniform rate region of any MAC with four users and binary inputs. & \textbf{Lemma 6.} $\mathcal{A}_4 \subset \mathrm{BMAT}_4 \subsetneq \mathrm{MAT}_4$. \\
\midrule
\addlinespace
\multicolumn{2}{@{}>{\centering\arraybackslash}p{\dimexpr0.96\textwidth+1.5em\relax}@{}}{%
  \itshape From ``A characterization of Johnson and Hamming graphs and proof of Babai's conjecture''
  (arXiv:1912.11427)%
} \\[2pt]
\textbf{Fact A.12.} Let $X$ be a generalized $2d$-gon of order $(s, s)$ for $2d \leq 6$, $s > 1$. Then the zero-weight spectral radius of $X$ satisfies $\xi(X) \leq 2s$. & \textbf{Theorem A.11.} A generalized $2d$-gon of order $(s, t)$ exists only for $2d \in \{4, 6, 8, 12\}$ unless $s = t = 1$. If $s > 1$, then $2d \neq 12$. \\
\end{longtable}
\endgroup

\begingroup
\small\setlength{\tabcolsep}{6pt}
\begin{longtable}{@{}p{0.48\linewidth}p{0.48\linewidth}@{}}
\caption{Notation edges the judge rejected. These show the dominant notation-extractor failure mode: shared symbols ($\Omega$, $\boldsymbol{S}$, $X:\Sigma\to\mathbb{S}^2\times\mathbb{R}$) without genuine dependency.}
\label{tab:dep-notation} \\
\toprule
\textbf{Source statement} & \textbf{Dependency statement} \\
\midrule
\endfirsthead
\caption[]{Notation edges the judge rejected (continued)} \\
\toprule
\textbf{Source statement} & \textbf{Dependency statement} \\
\midrule
\endhead
\bottomrule
\endfoot

\multicolumn{2}{@{}>{\centering\arraybackslash}p{\dimexpr0.96\textwidth+1.5em\relax}@{}}{%
  \itshape From ``Padé Approximants, density of rational functions in $A^\infty(\Omega)$\ldots''
  (arXiv:1212.4394)%
} \\[2pt]
\textbf{Proposition 6.3.} There exist a Jordan domain $\Omega$ and a function $f \in A(\Omega)$ such that the antiderivative of $f$ is not bounded inside $\Omega$. & \textbf{Lemma 5.4.} Let $\Omega$ be an open set, $\Omega \subseteq \mathbb{C}$. Then for every $N \in \mathbb{N}$ we have $\overline{\overline{(\Omega \cap D(0,N))}^{\,0}} = \overline{(\Omega \cap D(0,N))}$. \\
\midrule
\addlinespace
\multicolumn{2}{@{}>{\centering\arraybackslash}p{\dimexpr0.96\textwidth+1.5em\relax}@{}}{%
  \itshape From ``Weak Convergence of Probability Measures''
  (arXiv:2007.10293)%
} \\[2pt]
\textbf{Corollary 3.12 (Mapping theorem).} Let $h : \boldsymbol{S} \to \boldsymbol{S}'$ be a continuous mapping between two metric spaces. If $P_n \Rightarrow P$ on $\boldsymbol{S}$ and $P$ has a separable support, then $P_n h^{-1} \Rightarrow P h^{-1}$ on $\boldsymbol{S}'$. & \textbf{Theorem 3.9 (Prokhorov, direct part).} If a family of probability measures $\Pi$ on $(\boldsymbol{S}, \mathcal{S})$ is tight, then it is relatively compact. \\
\midrule
\addlinespace
\multicolumn{2}{@{}>{\centering\arraybackslash}p{\dimexpr0.96\textwidth+1.5em\relax}@{}}{%
  \itshape From ``The Gauss map of minimal surfaces in $\mathbb{S}^2 \times \mathbb{R}$''
  (arXiv:2006.09995)%
} \\[2pt]
\textbf{Proposition 4.7.} There is no minimal conformal immersion $X : \Sigma \to \mathbb{S}^2 \times \mathbb{R}$ whose Gauss map $g$ is an anti-holomorphic non-constant map. & \textbf{Proposition 4.3.} Let $X : \Sigma \to \mathbb{S}^2 \times \mathbb{R}$ be a minimal conformal immersion and $g$ its Gauss map. Then $g$ is constant if and only if $X(\Sigma)$ is part of a vertical cylinder over a geodesic of $\mathbb{S}^2$ in $\mathbb{S}^2 \times \mathbb{R}$. \\
\end{longtable}
\endgroup

\begingroup
\small\setlength{\tabcolsep}{6pt}
\begin{longtable}{@{}p{0.48\linewidth}p{0.48\linewidth}@{}}
\caption{Edges only the judge identified, typically genuine proof-level dependencies the parser missed because the author used the prior result implicitly.}
\label{tab:dep-judge-only} \\
\toprule
\textbf{Source statement} & \textbf{Dependency statement} \\
\midrule
\endfirsthead
\caption[]{Edges only the judge identified (continued)} \\
\toprule
\textbf{Source statement} & \textbf{Dependency statement} \\
\midrule
\endhead
\bottomrule
\endfoot

\multicolumn{2}{@{}>{\centering\arraybackslash}p{\dimexpr0.96\textwidth+1.5em\relax}@{}}{%
  \itshape From ``Graphs with large chromatic number induce $3k$-cycles''
  (arXiv:1408.2172)%
} \\[2pt]
\textbf{Theorem 8.} If $G$ is a connected trinity graph with large chromatic number and $r$ is a vertex of $G$, there exists a shadow or an antishadow. & \textbf{Theorem 1.} Every vertex $r$ in a connected trinity graph $G$ with large chromatic number is the origin of a large-order TCP. \\
\midrule
\addlinespace
\multicolumn{2}{@{}>{\centering\arraybackslash}p{\dimexpr0.96\textwidth+1.5em\relax}@{}}{%
  \itshape From ``Weak Convergence of Probability Measures''
  (arXiv:2007.10293)%
} \\[2pt]
\textbf{Theorem 1.6.} The following three conditions are equivalent: (i) $\boldsymbol{S}$ is separable; (ii) $\boldsymbol{S}$ has a countable base (a class of open sets such that each open set is a union of sets in the class); (iii) Each open cover of each subset of $\boldsymbol{S}$ has a countable subcover. & \textbf{Definition 1.5.} An \emph{open cover} of $A \subset \boldsymbol{S}$ is a class of open sets whose union contains $A$. \\
\midrule
\addlinespace
\multicolumn{2}{@{}>{\centering\arraybackslash}p{\dimexpr0.96\textwidth+1.5em\relax}@{}}{%
  \itshape From ``Extension of isometries between unit spheres of finite-dimensional polyhedral Banach spaces''
  (arXiv:1206.4839)%
} \\[2pt]
\textbf{Corollary 4.3.} If $\dim X = 2$, then $F$ is Gateaux differentiable at every non-zero point. & \textbf{Theorem 4.2.} In the following cases we can guarantee the Gateaux differentiability of $F$ at a point $y \in S_X$: (i) if $y \in \Sigma(X)$; (ii) if $\mathrm{lin}\, \gimel(y) = X^*$. \\
\end{longtable}
\endgroup
\newpage
\section{Retrieval and Matching Validation}
\label{sec:exp2}

This appendix provides the supporting data for the retrieval and judging claims made in \S\ref{sec:bridging}. We report the judge-calibration study behind our choice of judge of record, bidirectional open-pool retrieval on the blueprint pairs, the cross-project twin candidates surfaced by the same embedding, and the per-bin judging breakdown with representative judged examples.

\subsection{Choosing the judge}\label{app:calib}
Our first judge was Opus~4.8, run on all candidate pairs with a cosine similarity greater than $0.9$; a domain expert then sampled and graded ten of its judged matches. Table~\ref{tab:calib} shows these ten, each with its Opus~4.8 rating, Our expert's verdict and abridged note, and the rating GPT-5.4 produced on the same task. Our expert found Opus to be over-generous on five, over-calling matches as \texttt{exact} where the paper's result was only a special case or a different statement. The over-accepted cases recurred in three ways: the Lean statement was strictly more general than the paper's; a surface match hid a different underlying definition; or a more faithful Lean counterpart existed but was never surfaced.

These slip through because the judge works from compressed evidence. Sloganization abstracts away the hypotheses and definitions that separate a true match from a near one, so the judge over-indexes on the slogan. The Lean signature has the same problem from the other side: a generically named type can carry more specific assumptions than its name suggests, context an author or implementer reads off immediately but the judge never holds.

\begin{table*}[htbp]\centering\small
\begin{tabular}{@{}l c c c c p{6.8cm}@{}}\toprule
formal node & sim & Opus 4.8 & Expert & GPT-5.4 & Expert's note (abridged) \\\midrule
\texttt{append\_assoc} & 0.920 & \bx{green!20}{ex} & \bx{red!22}{wr} & \bx{yellow!35}{inx} & Neither rater grasps the graph structure: Lean's \texttt{SimpleGraph} is loopless (irreflexive), but the paper's graphs allow self-loops. A poor match. \\\addlinespace
\texttt{toMatrix\_adjoint} & 0.968 & \bx{green!20}{ex} & \bx{green!20}{ex} & \bx{yellow!35}{inx} & ``Given bases'' is imprecise from the slogan, but the paper means a unitary isomorphism (orthonormal basis), so it is exact, though not a faithful translation. \\\addlinespace
\texttt{lt\_of\_sub\_pos} & 0.909 & \bx{green!20}{ex} & \bx{green!20}{ex} & \bx{green!20}{ex} & Agree, up to the convention that $\mathbb{N}$ excludes $0$; too trivial to give any insight. \\\addlinespace
\texttt{inner\_tmul} & 0.927 & \bx{green!20}{ex} & \bx{red!22}{wr} & \bx{yellow!35}{inx} & Lean is wrong: Lean's is a general lemma on inner products of tensor products; the paper's is a special family of word spaces, an instance of Lean's theorem at best. \\\addlinespace
\texttt{measureEntropy} & 0.915 & \bx{yellow!35}{inx} & \bx{green!20}{ex} & \bx{yellow!35}{inx} & Outside my area, but Lean's looks slightly more general; likely a strong match. \\\addlinespace
\texttt{norm\_le\_interpStrip} & 0.928 & \bx{green!20}{ex} & \bx{red!22}{wr} & \bx{yellow!35}{inx} & Lean matches the standard Hadamard three-lines theorem, but the paper's printed Lemma~10 reverses the endpoint exponents, so as written it should be judged incorrect. \\\addlinespace
\texttt{iff\_rTensor} & 0.942 & \bx{green!20}{ex} & \bx{yellow!35}{inx} & \bx{yellow!35}{inx} & Exact after specializing to a commutative ring, but the paper's definition is for a left module over a possibly noncommutative ring while Mathlib is commutative-only. Inexact. \\\addlinespace
\texttt{coveringNumber} & 0.926 & \bx{green!20}{ex} & \bx{yellow!35}{inx} & \bx{red!22}{wr} & The informal is the usual \emph{external} covering number; Mathlib's is the \emph{internal} one (covering set $\subseteq A$). \texttt{externalCoveringNumber} is the better match. Inexact at best. \\\addlinespace
\texttt{exists\_prime} & 0.955 & \bx{green!20}{ex} & \bx{green!20}{ex} & \bx{green!20}{ex} & Agree, both are the natural-number form of Bertrand's postulate. An easy one. \\\addlinespace
\texttt{wnorm\_mono} & 0.945 & \bx{green!20}{ex} & \bx{green!20}{ex} & \bx{yellow!35}{inx} & Agree, monotonicity of the weak $L^p$ norm under a.e.\ domination; Lean is more general but a strong match. \\
\bottomrule
\end{tabular}
\caption{Expert calibration of ten Opus~4.8 matches at $\geq\!0.9$. For each candidate: its similarity and the verdicts of the original judge (Opus~4.8), the expert, and the judge of record (GPT-5.4), with the expert's abridged note. \bx{green!20}{ex}=exact, \bx{yellow!35}{inx}=inexact, \bx{red!22}{wr}=wrong. Opus matches the expert on four, over-calls \texttt{exact} on five, and is over-strict on one (\texttt{measureEntropy}); GPT-5.4 is stricter than Opus but, at this sample size, does not fully track the expert.}
\label{tab:calib}
\end{table*}

\subsection{Calibrating the judge of record}\label{app:calib-gpt}
Having rejected Opus, we now validate the judge of record. The same domain expert graded a fresh ten candidates spanning the full $\geq\!0.8$ band, with GPT-5.4's verdict visible only after the expert's own call. Table~\ref{tab:calib-gpt} shows the result.

\begin{table*}[htbp]\centering\small
\begin{tabular}{@{}l c c c p{6.8cm}@{}}\toprule
formal node & sim & Expert & GPT-5.4 & Expert's note (abridged) \\\midrule
\texttt{rpow\_add\_le\_mul\_rpow\_add\_rpow} & 0.974 & \bx{green!20}{ex} & \bx{green!20}{ex} & Same inequality. Lean's \texttt{NNReal} is $[0,\infty)$, potentially slightly stronger than the paper if its $\mathbb{R}_+$ is open. \\\addlinespace
\texttt{PerfectlyNormalSpace} & 0.957 & \bx{green!20}{ex} & \bx{green!20}{ex} & Same definition. The paper's standing ``all spaces are T1'' assumption is slightly stronger than Mathlib's \texttt{PerfectlyNormalSpace}; T1 is bundled separately in \texttt{T6Space}. \\\addlinespace
\texttt{LieSubalgebra.lie\_mem} & 0.927 & \bx{yellow!35}{inx} & \bx{yellow!35}{inx} & Lean is a field accessor on an existing \texttt{LieSubalgebra}; the informal asserts the definition plus the inheritance claim that $\mathfrak{h}$ is itself a Lie algebra. \\\addlinespace
\texttt{alternatingGroup\_le\_\ldots\_isThreeCycle\_mem} & 0.917 & \bx{green!20}{ex} & \bx{green!20}{ex} & Jordan's theorem in both. \texttt{IsPreprimitive} differs from classical primitivity only in nonemptiness, forced here by the 3-cycle; faithfulness is automatic from $G \le \mathrm{Equiv.Perm}\,\alpha$. \\\addlinespace
\texttt{AEMeasurable.lintegral\_prod\_right'} & 0.914 & \bx{red!22}{wr} & \bx{red!22}{wr} & Lean is a.e.-measurability of the inner lower integral on $\mathbb{R}_{\ge 0}^\infty$; informal is the full vector-valued Bochner Fubini theorem. \\\addlinespace
\texttt{Measure.finiteAt\_nhdsWithin} & 0.882 & \bx{red!22}{wr} & \bx{yellow!35}{inx} & Informal is the \emph{definition} of locally finite. Lean is a one-line filter-level consequence assuming it; the faithful counterpart is the class \texttt{IsLocallyFiniteMeasure}. \\\addlinespace
\texttt{HasDerivAt.inner} & 0.880 & \bx{yellow!35}{inx} & \bx{yellow!35}{inx} & Special case of the informal directional-derivative product rule, restricted to ordinary derivatives on $\mathbb{R} \to E$. \\\addlinespace
\texttt{starIsoTerminal\_inv} & 0.865 & \bx{red!22}{wr} & \bx{red!22}{wr} & Informal is the definition of a terminal object. Lean is an equality between two specific morphisms inside the constructed category \texttt{WithTerminal\,C}; uses terminal-object machinery without stating the universal property. \\\addlinespace
\texttt{add\_lt\_add\_of\_le\_of\_lt} & 0.844 & \bx{red!22}{wr} & \bx{red!22}{wr} & Paper is non-strict on both sides; Lean takes mixed non-strict/strict and concludes a strict inequality. The closer counterpart is \texttt{add\_le\_add}. \\\addlinespace
\texttt{map\_neighborFinset\_induce} & 0.822 & \bx{yellow!35}{inx} & \bx{red!22}{wr} & Same structural idea but the paper is directed in-neighborhoods in a sub-tournament; Lean is undirected \texttt{SimpleGraph} neighbors. A faithful cross-category analogue, not the same statement. \\
\bottomrule
\end{tabular}
\caption{Calibration of GPT-5.4 against the expert on ten candidates spanning the full $\geq\!0.8$ band. \bx{green!20}{ex}=exact, \bx{yellow!35}{inx}=inexact, \bx{red!22}{wr}=wrong. GPT-5.4 agrees with the expert on $8/10$ verdicts strictly, and on $9/10$ when collapsed to match/non-match (\texttt{exact}${+}$\texttt{inexact} vs.\ \texttt{wrong}).}
\label{tab:calib-gpt}
\end{table*}

The two disagreements are scope-direction errors in opposite directions. On \texttt{Measure.finiteAt\_nhdsWithin}, the informal is the definition of local finiteness and the Lean lemma is a one-line consequence assuming it; the expert reads this as a level mismatch and grades \texttt{wrong}, GPT-5.4 softens to \texttt{inexact}. On \texttt{map\_neighborFinset\_induce}, the Lean lemma is the same structural statement as the paper's claim transposed from directed graphs to undirected; the expert reads this as a faithful cross-category analogue and grades \texttt{inexact}, GPT-5.4 hardens to \texttt{wrong}.

On the eight agreed verdicts, the expert's and judge's rationales tend to name the same distinction in different words. At the strict-vs-non-strict level (\texttt{add\_lt\_add\_of\_le\_of\_lt}) and at the integration-theory level (\texttt{lintegral\_prod\_right'}) GPT-5.4 catches the gap without prompting. We therefore retain GPT-5.4 as the judge of record, and read the per-bin totals in Table~\ref{tab:judge-breakdown} with the understanding that the strict \texttt{wrong}/\texttt{inexact} split is judge-dependent in the way this calibration documents.

\subsection{Open-pool retrieval}
We also validated retrieval bidirectionally on the blueprint pairs.\footnote{Blueprint pairs are the reference standard throughout this appendix. Confidence intervals on the retrieval and twin-count numbers are $2{,}000$-resample bootstraps (seed~42).} We queried each pair's endpoint and checked whether its annotated partner was returned at rank~1 (Table~\ref{tab:bp-retrieval}). The two directions search different candidate sets. In the formal-to-informal direction ($f\to i$), the formal statement queries the full 11.7M-statement informal corpus, exactly as in the discovery sweep of \S\ref{sec:bridging}. In the informal-to-formal direction ($i\to f$), the informal statement queries only the $36{,}708$ declarations from the formalization projects, not the full \nSwept-declaration formal graph searched in \S\ref{sec:bridging}. This restriction is appropriate because every blueprint \texttt{\textbackslash lean\{\}} target is a project declaration; the remaining Mathlib and Lean-core declarations contain no possible answer and would only add distractors.

We find that disjoint pairs are largely missed across both directions of querying. In particular, $61.2\%$ $[58.9, 63.6]$ of blueprint pairs are recovered by at least one direction, and $22.3\%$ $[20.2, 24.4]$ are mutual rank-1. As a non-semantic baseline, we also reran $i\to f$ over the same project declarations using BM25, a standard lexical ranker based solely on weighted query--candidate term overlap. Because the embedding-based method outperforms BM25 at every cutoff, its matches cannot be explained solely by shared surface vocabulary and instead reflect semantic agreement.

\begin{table}[htbp]
  \centering \small \setlength{\tabcolsep}{5pt}
  \begin{tabular}{lrrrr}
    \toprule
    Sweep & Hit@1 & Hit@5 & Hit@10 & MRR \\
    \midrule
    Embedding $f\to i$  & \textbf{43.5} & 64.8 & 69.9 & 52.5 \\
    Embedding $i\to f$  & \textbf{42.6} & 67.0 & 71.7 & 53.1 \\
    BM25 $i\to f$       & 31.3          & 53.7 & 61.0 & 41.0 \\
    \bottomrule
  \end{tabular}
  \caption{Open-pool retrieval (\%). $n$ is the number of distinct evaluable queries: $1{,}530$ of the $1{,}577$ blueprint formals for $f\to i$ (47 lack an evaluable open-pool partner), $1{,}308$ blueprint informals for $i\to f$. BM25 is a lexical baseline.}
  \label{tab:bp-retrieval}
\end{table}

\subsection{Match types}
\begin{figure}[htbp]
  \centering
  \includegraphics[width=\linewidth]{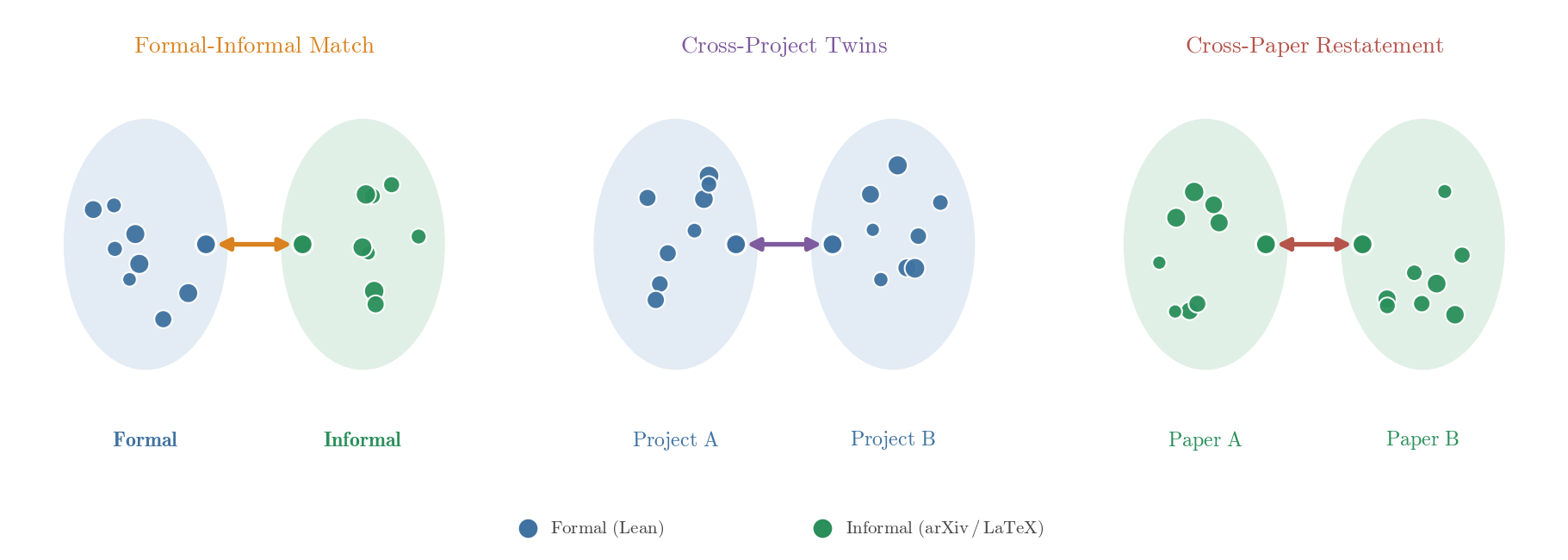}
  \caption{Three kinds of match the slogan embedding can surface, by the corpora of the query and its nearest neighbor: a formal--informal match ($f\!\leftrightarrow\!i$), a cross-project twin ($f\!\leftrightarrow\!f$), and a cross-paper restatement ($i\!\leftrightarrow\!i$).}
  \label{fig:match-types}
\end{figure}

Because both graphs are embedded in one space, the same slogan retrieval surfaces three kinds of match depending on which corpora the query and its neighbor come from (Figure~\ref{fig:match-types}): a \emph{formal--informal match} ($f\!\leftrightarrow\!i$), the matches of \S\ref{sec:bridging} which the open-pool retrieval above validates; a \emph{cross-project twin} ($f\!\leftrightarrow\!f$), the same result formalized in two projects, which we probe below; and a \emph{cross-paper restatement} ($i\!\leftrightarrow\!i$), the same result stated in two papers, which we do not pursue here.

\subsection{Candidate cross-project twins ($f\to f$)}
The same slogan embedding can be swept formal-to-formal across the project declarations to surface \emph{candidate} twins: a project lemma whose nearest neighbor is a lemma in a different project. Excluding same-paper matches and two parallel-formalization repository pairs, $446$ $[253, 682]$ pairs sit at cosine $\geq 0.85$ (14 at $\geq 0.95$); Table~\ref{tab:twins-examples} lists the highest-cosine ones. Consistent with \S\ref{sec:bridging}, cosine here is a candidate signal, not ground truth: these pairs are \emph{not} judge-confirmed, so the count is a candidate set rather than a verified twin set. Running the same LLM-as-judge step over these candidates to confirm genuine twins is a natural extension we have not yet performed.

\begin{table}[htbp]
  \centering \small \setlength{\tabcolsep}{4pt}
  \begin{tabular}{rp{0.74\linewidth}}
    \toprule
    sim & Candidate twin pair and shared statement \\
    \midrule
    0.999 & \texttt{NNReal.coe\_comp\_dconv} (add-combi) $\leftrightarrow$ \texttt{NNReal.coe\_comp\_cdconv} (apap): the $\mathbb{R}_{\ge 0}\to\mathbb{R}$ coercion commutes with discrete convolution. \\
    \addlinespace
    0.988 & \texttt{$\Phi_4'$\_contDiffOn} (Sphere-Packing-Lean) $\leftrightarrow$ same (sphere-packing-math-inc): $\Phi_4'$ is continuously differentiable on its domain. \\
    \addlinespace
    0.987 & \texttt{neg\_cdconv} (apap) $\leftrightarrow$ \texttt{neg\_conv} (add-combi): negation distributes through the convolution. \\
    \bottomrule
  \end{tabular}
  \caption{Highest-cosine $f\to f$ candidate twins. The top pairs are likely equivalent statements under independent identifier conventions, but the set as a whole is cosine-surfaced and not judge-confirmed.}
  \label{tab:twins-examples}
\end{table}

\subsection{Judging breakdown}
Table~\ref{tab:judge-breakdown} gives the per-bin verdict counts behind Figure~\ref{fig:comp}: for each cosine band, how many judged candidates each model labels \texttt{exact}, \texttt{inexact}, or \texttt{wrong}, and the resulting match rate (\texttt{exact}${+}$\texttt{inexact}). Both judges move from almost all matches at the top of the band, a mix of \texttt{exact} and \texttt{inexact}, to mostly \texttt{wrong} near the floor; DeepSeek is uniformly more generous. The two models judge near-identical candidate sets, with small per-bin differences in $n$ from parse failures and unjudgeable cases.

\begin{table}[htbp]
  \centering \small \setlength{\tabcolsep}{5pt}
  \begin{tabular}{lrrrrr}
    \toprule
    Sim.\ bin & $n$ & \bx{green!20}{exact} & \bx{yellow!35}{inexact} & \bx{red!22}{wrong} & match (\%) \\
    \midrule
    \multicolumn{6}{@{}c}{\itshape GPT-5.4 (judge of record)}\\
    0.95--1.0  &     676 &     290 &     360 &      26 &    650 (96.2) \\
    0.90--0.95 &  6{,}644 & 1{,}391 &  4{,}312 &     941 &  5{,}703 (85.8) \\
    0.85--0.90 & 26{,}755 & 1{,}834 & 14{,}852 & 10{,}069 & 16{,}686 (62.4) \\
    0.80--0.85 & 66{,}537 & 1{,}250 & 23{,}663 & 41{,}624 & 24{,}913 (37.4) \\
    \addlinespace
    \textbf{Total} & 100{,}612 & 4{,}765 & 43{,}187 & 52{,}660 & \textbf{47{,}952 (47.7)} \\
    \midrule
    \multicolumn{6}{@{}c}{\itshape DeepSeek-V4-Pro (lenient comparison)}\\
    0.95--1.0  &     678 &     587 &      80 &      11 &    667 (98.4) \\
    0.90--0.95 &  6{,}645 &  4{,}542 &  1{,}678 &     425 &  6{,}220 (93.6) \\
    0.85--0.90 & 26{,}827 & 11{,}151 &  9{,}195 &  6{,}481 & 20{,}346 (75.8) \\
    0.80--0.85 & 66{,}700 & 14{,}951 & 19{,}050 & 32{,}699 & 34{,}001 (51.0) \\
    \addlinespace
    \textbf{Total} & 100{,}850 & 31{,}231 & 30{,}003 & 39{,}616 & \textbf{61{,}234 (60.7)} \\
    \bottomrule
  \end{tabular}
  \caption{Per-bin judging breakdown for the two judges over the corrected candidate pool ($\geq\!0.8$ cosine), highest band first. Column headers are colored by verdict (\bx{green!20}{exact}, \bx{yellow!35}{inexact}, \bx{red!22}{wrong}); \texttt{exact}${+}$\texttt{inexact} count as matches. GPT-5.4 is our set of record, DeepSeek-V4-Pro a more lenient high-recall comparison; the bold totals contrast the strict (47.7\%) and lenient (60.7\%) overall match rates.}
  \label{tab:judge-breakdown}
\end{table}

\subsection{Judged match examples}
\label{app:matching-examples}
Tables~\ref{tab:matching-verdict} and~\ref{tab:matching-band} show representative GPT-5.4-judged candidate edges from the corrected set, each pairing the formal Lean statement with its informal arXiv counterpart. Table~\ref{tab:matching-verdict} gives one example per verdict class so the difference between a match and a false positive is visible; Table~\ref{tab:matching-band} gives judge-affirmed (\texttt{exact}) matches at a high, mid, and low point of the $0.9$--$1.0$ band.

\begin{table*}[htbp]\centering\small\setlength{\tabcolsep}{6pt}
\begin{tabular}{@{}>{\raggedright\arraybackslash}p{\dimexpr0.5\textwidth-2.8cm\relax} >{\raggedright\arraybackslash}p{\dimexpr0.5\textwidth-2.8cm\relax} >{\footnotesize\raggedright\arraybackslash}p{1.7cm} c c@{}}
\toprule
\textbf{Formal (Lean)} & \textbf{Informal statement} & \multicolumn{1}{l}{\textbf{arXiv}} & sim & verdict \\
\midrule
\lc{niven \{$\theta$ : $\mathbb{R}$\} (h$\theta$ : $\exists$ r, $\theta$ = $\uparrow$r * Real.pi) (hcos : $\exists$ q, Real.cos $\theta$ = $\uparrow$q) : Real.cos $\theta$ $\in$ \{-1, -1 / 2, 0, 1 / 2, 1\}} & Let $\theta$ be a rational multiple of $2\pi$. If $\cos(\theta)\in\mathbb{Q}$, then $\cos(\theta)\in\left\{-1,-\tfrac12,0,\tfrac12,1\right\}$. & 1410.6912 & 0.97 & \bx{green!20}{exact} \\
\addlinespace
\lc{SemidirectProduct.mul\_def \{N : Type u\_1\} \{G : Type u\_2\} [Group N] [Group G] \{$\varphi$ : G $\to$* MulAut N\} (a b : N $\rtimes$[$\varphi$] G) : a * b = $\langle$a.left * ($\varphi$ a.right) b.left, a.right * b.right$\rangle$} & The semidirect product $N\rtimes_\varphi H$ has underlying set $N\times H$ with $(n_1,h_1)(n_2,h_2)=(n_1\varphi(h_1)(n_2),h_1h_2)$. & 2111.00575 & 0.90 & \bx{yellow!35}{inexact} \\
\addlinespace
\lc{Ne.symm \{$\alpha$ : Sort u\} \{a b : $\alpha$\} (h : a $\neq$ b) : b $\neq$ a} & For all $a,b$, if $a=b$ then $b=a$. & 0707.1981 & 0.90 & \bx{red!22}{wrong} \\
\bottomrule
\end{tabular}
\caption{Representative GPT-5.4 verdicts from the corrected candidate set, one per class: the verbatim Lean statement, the paper's informal statement, the arXiv id, the cosine similarity, and the verdict. \texttt{exact}/\texttt{inexact} count as matches; \texttt{wrong} (lexically near but mathematically distinct, e.g.\ \texttt{Ne.symm}: symmetry of $\neq$ vs.\ the paper's $=$) does not.}
\label{tab:matching-verdict}
\end{table*}

\begin{table*}[htbp]\centering\small\setlength{\tabcolsep}{6pt}
\begin{tabular}{@{}>{\raggedright\arraybackslash}p{\dimexpr0.5\textwidth-2.8cm\relax} >{\raggedright\arraybackslash}p{\dimexpr0.5\textwidth-2.8cm\relax} >{\footnotesize\raggedright\arraybackslash}p{1.7cm} c c@{}}
\toprule
\textbf{Formal (Lean)} & \textbf{Informal statement} & \multicolumn{1}{l}{\textbf{arXiv}} & sim & verdict \\
\midrule
\lc{Aux.sum\_inv\_le\_log (n : $\mathbb{N}$) (hn : 1 $\le$ n) : $\sum$ d $\in$ Finset.Icc 1 n, ($\uparrow$d)$^{-1}$ $\le$ 1 + Real.log $\uparrow$n} & For any positive integer $n$, $\sum_{t=1}^{n}\tfrac{1}{t}\le 1+\log n$. & 2002.05359 & 0.98 & \bx{green!20}{exact} \\
\addlinespace
\lc{Real.cos\_sq\_add\_sin\_sq (x : $\mathbb{R}$) : Real.cos x \textasciicircum{} 2 + Real.sin x \textasciicircum{} 2 = 1} & The identity $\sin^2 x+\cos^2 x=1$ holds for $x\in\mathbb{R}$. & 2312.13525 & 0.95 & \bx{green!20}{exact} \\
\addlinespace
\lc{legendreSym.at\_neg\_two \{p : $\mathbb{N}$\} [Fact (Nat.Prime p)] (hp : p $\neq$ 2) : legendreSym p (-2) = ZMod.$\chi_8'$ $\uparrow$p} & If $p$ is an odd prime, $\left(\tfrac{-2}{p}\right)=1$ if $p\equiv 1,3 \pmod 8$ and $-1$ if $p\equiv 5,7 \pmod 8$. & 2309.02512 & 0.91 & \bx{green!20}{exact} \\
\bottomrule
\end{tabular}
\caption{Judge-affirmed (\texttt{exact}) matches from the corrected GPT-5.4 set at a high, mid, and low point of the $0.9$--$1.0$ band (similarity $\approx$ 0.98 / 0.95 / 0.91): the verbatim Lean statement against the paper's informal statement. \texttt{legendreSym.at\_neg\_two} shows the abstract Lean form ($=\chi_8'(p)$) the judge matched to the paper's explicit residue cases.}
\label{tab:matching-band}
\end{table*}
\newpage
\section{Corpus Composition by Library}
\label{app:corpus}

Our dependency graph spans 388{,}105 formal declarations and 9{,}585{,}510 within-library dependency edges across 25 Lean libraries. Mathlib dominates both, supplying 90.5\% of declarations and 97.4\% of within-library edges; the Mathlib rows aggregate the v4.27--v4.29 snapshots. The remaining libraries are research/blueprint developments, each contributing at most a few thousand declarations. Their internal edge counts are sparser still: most of their dependencies point into Mathlib, and such cross-library edges are excluded here---each count reflects only edges whose source and target lie in the same library. Table~\ref{tab:corpus} reports declarations and within-library edges side by side.

\begin{table}[H]
\centering\scriptsize
\setlength{\tabcolsep}{3pt}
\begin{tabular}{lrlr}
\toprule
\multicolumn{2}{c}{\textbf{Declarations}} & \multicolumn{2}{c}{\textbf{Within-lib.\ edges}} \\
\cmidrule(r){1-2}\cmidrule(l){3-4}
\textbf{Library} & \textbf{Count} & \textbf{Library} & \textbf{Count} \\
\midrule
Mathlib (v4.27--v4.29) & 351{,}397 & Mathlib (v4.27--v4.29) & 9{,}333{,}251 \\
\midrule
\texttt{physlib} & 8{,}205 & \texttt{physlib} & 93{,}688 \\
\texttt{PrimeNumberTheoremAnd} & 5{,}108 & \texttt{carleson} & 27{,}553 \\
\texttt{sphere-packing-math-inc} & 4{,}125 & \texttt{sphere-packing-math-inc} & 25{,}867 \\
\texttt{carleson} & 2{,}852 & \texttt{combinatorial-games} & 24{,}666 \\
\texttt{combinatorial-games} & 2{,}702 & \texttt{PrimeNumberTheoremAnd} & 20{,}818 \\
\texttt{FLT} & 2{,}368 & \texttt{cslib} & 15{,}042 \\
\texttt{cslib} & 2{,}273 & \texttt{FLT} & 10{,}833 \\
\texttt{ClassFieldTheory} & 1{,}387 & \texttt{sphere-eversion} & 6{,}740 \\
\texttt{sphere-eversion} & 1{,}208 & \texttt{pfr} & 5{,}671 \\
\texttt{pfr} & 1{,}073 & \texttt{ClassFieldTheory} & 4{,}482 \\
\texttt{brownian-motion} & 1{,}071 & \texttt{brownian-motion} & 3{,}124 \\
\texttt{Sphere-Packing-Lean} & 821 & \texttt{toric} & 2{,}630 \\
\texttt{apap} & 670 & \texttt{HarderNarasimhan} & 2{,}320 \\
\texttt{formal-conjectures} & 639 & \texttt{apap} & 2{,}155 \\
\texttt{toric} & 485 & \texttt{flt-regular} & 1{,}361 \\
\texttt{misc-yd} & 362 & \texttt{misc-yd} & 1{,}200 \\
\texttt{flt-regular} & 344 & \texttt{formal-conjectures} & 976 \\
\texttt{HarderNarasimhan} & 309 & \texttt{Sphere-Packing-Lean} & 945 \\
\texttt{add-combi} & 190 & \texttt{PersistentDecomp} & 785 \\
\texttt{PersistentDecomp} & 161 & \texttt{add-combi} & 507 \\
\texttt{cam-combi} & 132 & \texttt{gibbs-measure} & 341 \\
\texttt{gibbs-measure} & 113 & \texttt{forbidden-matrix} & 308 \\
\texttt{forbidden-matrix} & 86 & \texttt{cam-combi} & 170 \\
\texttt{chandra-furst-lipton} & 24 & \texttt{chandra-furst-lipton} & 77 \\
\midrule
\textbf{Total} & \textbf{388{,}105} & \textbf{Total} & \textbf{9{,}585{,}510} \\
\bottomrule
\end{tabular}
\caption{Corpus composition by library: declarations and within-library dependency edges. Cross-library edges are excluded; non-Mathlib libraries contribute 36{,}708 declarations (9.5\%) and 252{,}259 within-library edges (2.6\%).}
\label{tab:corpus}
\end{table}
\newpage
\section{Extending the Corpus Beyond arXiv: PDF Ingestion}
\label{app:pdf-ingestion}

The informal graph (the \emph{Informal Graph} section) is built entirely from arXiv \LaTeX{} source, which excludes a large body of mathematical writing---textbooks, lecture notes, and older papers---that exists only as PDF. As a first step toward extending the corpus beyond arXiv, we built a PDF ingestion tool that recovers statement-level content from PDF sources in the same schema as our existing non-arXiv ingests (e.g.\ ProofWiki). We describe the pipeline and report a single-source run on Artin's \emph{Algebra} \cite{artin2011algebra}; we treat this as a feasibility study rather than a contribution to the headline corpus, since the resulting statements lack the cross-reference structure the deterministic and heuristic extractors rely on.

\paragraph{Two-stage extraction.}
Optical character recognition over an entire textbook is slow and largely wasted, since most pages carry no theorem-like content. We therefore split ingestion into a cheap localization pass and an expensive recognition pass. A \texttt{pymupdf} \cite{pymupdf2026} text scan first identifies pages whose text contains a numbered environment header (\texttt{Theorem}, \texttt{Lemma}, \texttt{Proposition}, \texttt{Corollary}, \texttt{Definition}, and similar) at the start of a line, distinguishing genuine headers from in-text references such as ``by Theorem~1.2.'' Each matched page is retained together with a one-page buffer on either side, so that proofs spilling onto an adjacent page are not truncated. The retained pages are copied into a temporary PDF, and only that reduced document is passed to Nougat OCR \cite{nougat-ocr}, which emits Mathpix Markdown (\texttt{.mmd}): a Markdown-style format with mathematics in \LaTeX{} delimiters \cite{mathpix_markdown}. On Artin's \emph{Algebra}, 182 of 555 pages carry a numbered environment header, so OCR runs on roughly a third of the book plus buffer pages rather than the whole volume.

\paragraph{Statement parsing.}
The Markdown output is parsed into statement records by a line-oriented pass that recognizes both Nougat's bold-header style (\verb|**Theorem 1.2**|) and plain numbered headers. For each environment it records the type, the environment number, an optional inline title, the statement body, and---when present---the proof, which is delimited by a leading \emph{Proof} marker and a closing \texttt{QED} symbol ($\square$ or $\blacksquare$). Markdown emphasis is stripped while inline and display math spans are masked and restored intact, so the recognized \LaTeX{} survives the cleaning step. The output is a JSON list of statement records matching the schema of our other non-arXiv sources, which lets ingested PDFs enter the corpus through the same path as ProofWiki without a separate \LaTeX{}-source stage. Multiple PDFs placed in the input directory are processed in a single batch.

\paragraph{Feasibility run.}
Table~\ref{tab:pdf-artin} reports the result of running the pipeline on Artin's \emph{Algebra}. The tool recovers 406 statements, 242 of which (59.6\%) carry an extracted proof, and every recovered statement retains its environment number. The recovered counts track---and slightly exceed---a strict line-start header count of the source PDF (75 theorems, 104 propositions, 59 lemmas, 55 corollaries, 13 numbered definitions), indicating that the parser captures essentially all numbered environments plus a small number formatted irregularly.

\begin{table}[H]
\centering\small
\setlength{\tabcolsep}{6pt}
\begin{tabular}{lrr}
\toprule
\textbf{Type} & \textbf{Recovered} & \textbf{With proof} \\
\midrule
Proposition & 126 & 91 \\
Theorem     &  92 & 52 \\
Lemma       &  81 & 64 \\
Corollary   &  66 & 34 \\
Example     &  26 &  1 \\
Definition  &  15 &  0 \\
\midrule
\textbf{Total} & \textbf{406} & \textbf{242} \\
\bottomrule
\end{tabular}
\caption{PDF ingestion of Artin's \emph{Algebra}. ``With proof'' counts statements for which a delimited proof block was extracted. Proof-bearing counts are illustrative of the proof-splitting behavior and were not separately validated.}
\label{tab:pdf-artin}
\end{table}

\paragraph{Definition yield and authorial style.}
The low definition count (15) is not an extraction failure but a property of the source. The book contains only 13 numbered \texttt{Definition} environments; across 102 occurrences of the word ``definition'' in the text, the remainder are prose definitions introduced inline and referred back to throughout (``the definition of matrix multiplication,'' ``a recursive definition of the determinant''). Artin formalizes a small definitional core as numbered environments and reuses it in running prose, so definition yield reflects how a given text sets off its definitions rather than the recognizer's coverage. This source-dependence is a general feature of OCR-based ingestion: extraction recovers what a text marks as a numbered environment, and the proportion of content exposed this way varies by author and by subject area.

\paragraph{Limitations.}
PDF-derived statements arrive without the machine-readable cross-reference structure that drives the informal extractors. There are no \verb|\label| keys, no \verb|\ref| commands, and no resolvable reference list, so the deterministic extractor---which achieves 98.8\% judge-verified precision on arXiv source---cannot propose within-source edges for these statements, and the heuristic extractor is limited to prose cues such as ``Theorem~3.2.'' Dependency recovery for PDF sources therefore rests largely on the notation extractor and on the semantic restatement edges of the universal graph as seen in Section~\ref{sec:universal-graph}. Bringing PDF sources to parity with arXiv ingestion---in particular, recovering usable within-source references from OCR'd text---is left to future work.
\newpage
\section{Improving Premise Selection of Theorem-Proving Agents}
\subsection{Introduction}
Automated theorem proving and AI-assisted mathematical reasoning depend on premise selection: given a target theorem, the system must find earlier lemmas and theorems that can be used in its proof. The quality of these premises largely determines the size of the proof search problem. A relevant lemma can reduce a difficult goal to a known argument; an irrelevant neighbor increases the number of proof paths the system must consider. The main difficulty is that semantic similarity is only a proxy for logical dependence. A theorem is often close in embedding space to statements about the same objects, notation, or topic. Its proof, however, may cite earlier results that introduce a tool, a reduction, a classification theorem, or a structural fact whose statement is semantically distinct from the target. A semantic search system can therefore retrieve plausible mathematical neighbors while missing the premises that a proof actually uses.

This experiment examines premise selection in a custom environment that wraps TheoremSearch \cite{theoremsearch2026}. TheoremSearch ranks statements by similarity to a natural-language query, but it does not itself learn which query phrasings are likely to retrieve dependencies. We frame query generation as a sequential decision problem: Given a target theorem, an agent issues a fixed number of natural-language queries, observes the results, and is rewarded for retrieving true dependencies.

This motivates the question of whether reinforcement learning can teach a language model to use semantic search in a way that respects dependency structure. Our results are mixed: training improves the base Qwen3-8B policy, but performance plateaus below that of frontier models such as Gemini 3.1 Pro and GPT-5.5. Inspection of the learned queries indicates that the policy can formulate coherent, mathematically precise searches but generally lacks the detailed knowledge of the literature required to recognize and retrieve the cited result. A simple expansion over the citation graph recovers substantially more dependencies, suggesting that much of the remaining signal is encoded in the corpus’s graph structure rather than in embedding-based similarity alone.

\subsection{Methodology}
We formulate premise selection as an MDP over the TheoremSearch API. The state contains the target theorem, represented by its natural-language slogan and \LaTeX\ body, together with the set of statements returned by previous queries. The action is a query string passed to a TheoremSearch API call, which returns the top-$k$ statements by cosine similarity over Qwen3-Embedding-8B slogan embeddings \cite{zhang2025qwen3embedding}. The agent may issue at most $H$ queries. The reward gives $+1$ for each newly retrieved true dependency, applies a false-positive penalty $-\alpha$, and gives a terminal bonus proportional to final dependency recall.

The policy is a Qwen3 language model trained to generate query strings conditioned on the current state. We evaluate dependency recall at $k{=}10$ over $H{=}6$ queries. We use the same configuration for the prompted GPT-5.5 and Gemini-3.1-Pro baselines, so comparisons isolate the effect of the query policy rather than the retrieval budget. At each turn, the policy emits a short reasoning trace before the query. This matches the prompted baselines and, empirically, accounts for a substantial part of the improvement over the base model.

\paragraph{Self-distillation objective.} The retrieval reward is sparse and non-differentiable: most sampled query sequences retrieve no true dependencies, and the reward cannot directly supervise individual query tokens. We therefore train the agent using Self-Distillation Policy Optimization (SDPO) \cite{hubotter2026sdpo}, which uses the best rollout from a sampled group as a feedback signal for policy improvement. For each target theorem, we sample $G{=}4$ rollouts. We select the rollout with the highest dependency recall and use its query strings as feedback $f$. The student policy is then trained to match a feedback-conditioned teacher policy without receiving that feedback at inference time. The per-token loss is a symmetric Jensen--Shannon divergence over the generated completion:
\[
	\mathcal{L}_{\text{SDPO}}(\theta)=\sum_t \mathrm{JS}\!\left(\pi_\theta(\cdot\mid x,y_{<t})\,\|\,
	\mathrm{sg}\,\pi_{\text{teacher}}(\cdot\mid x,f,y_{<t})\right).
\]
The teacher is an exponential moving average of the LoRA weights with $\alpha_{\text{EMA}}{=}0.01$, and $\mathrm{sg}$ denotes stop-gradient. The feedback is the agent's own highest-recall query sequence from the sampled group, rather than the ground-truth dependency list. We additionally evaluate a hybrid SDPO+GRPO objective. SDPO trains the policy to imitate the
highest-reward rollout within a sampled group, while GRPO directly optimizes expected reward using group-relative advantages. The hybrid objective optimizes the sum of the SDPO imitation loss and the GRPO policy-gradient loss:
\[
L_{\text{hybrid}}(\theta)
=
L_{\text{SDPO}}(\theta)
+
\lambda L_{\text{GRPO}}(\theta),
\]
with $\lambda \in [0,1]$. For our test, we used $\lambda = 0.5$.

\paragraph{Training and evaluation context.} During sampled training rollouts, the agent sees a compact rendering of each retrieved statement: the slogan (a natural-language summary of the statement), the paper title, the first author, and the citation count. This lets the policy condition later queries on earlier results. However, in greedy evaluation, including retrieved results lowers recall. The longer context shifts the model toward commenting on previous outputs rather than producing a sharper next query. The final reported policy is therefore trained with result context but evaluated without it. All trained policies are LoRA adapters of rank 16 on Qwen3-8B, which update about 0.53\% of the model parameters.

\paragraph{Citation-graph expansion.} TheoremSearch can also return a statement's one-hop dependency neighborhood; i.e., all results that are one dependency edge away from the target. Since premise selection is defined by citation-derived dependencies, we test whether retrieved statements can serve as seeds for graph expansion. After the policy retrieves statements with embedding search, we add the cross-paper one-hop graph neighbors of those statements to the candidate set. We never expand the target statement itself. To prevent data leakage, we also remove any statement from the target's own paper before computing recall. This prevents the evaluation from rewarding a shortcut in which the system copies citations from sibling theorems in the same paper.

\subsection{Related work}
Premise selection is usually studied inside proof-assistant libraries, where a retriever supplies candidate premises to a prover. LeanDojo retrieves premises for Lean proofs \cite{leandojo}, LeanSearch provides semantic search over Mathlib \cite{leansearch-v1, leansearch-v2}, and earlier work trains language models to generate proofs directly \cite{polu2020generative}. This project differs in two ways. First, retrieval itself is the task: the agent is evaluated on whether it recovers dependency edges, not on whether a downstream prover succeeds. Second, the search space is a broad mathematical corpus indexed by dense slogan embeddings, closer to dense-passage retrieval \cite{karpukhin2020dpr,zhang2025qwen3embedding} than to a single formal library.

The learning setup is related to RL-based agentic search, in which a language model learns to interact with a search environment through multi-step interactions \cite{lin2025agenticsearch}. We train with SDPO as the objective \cite{hubotter2026sdpo} because it has been shown to improve sample efficiency and accuracy over strong RLVR baselines \cite{hubotter2026sdpo}. The graph-augmentation experiment is also related to work showing that language models can benefit from explicit graph structure \cite{guo2025g1,zhang2025generalizable}. In this report, the graph is not learned by the model; it is used at retrieval time to expose dependency edges adjacent to the statements the model already retrieves.

\subsection{Results}
\label{sec:results}

\begin{table}[h]
	\centering
	\caption{Dependency recall at ($k{=}10$, $H{=}6$).}
	\label{tab:main}
	\begin{tabular}{lccl}
		\toprule
		System                                         & Recall         & Mean \# of retrieved candidates                     \\
		\midrule
		Qwen3-8B base                                  & $\sim$0.04    & $\sim$60     \\
		Qwen3-8B SDPO                          & 0.079          & $\sim$60  \\
		\quad + cross-paper graph augmentation\  & \textbf{0.134} & $\sim$115 \\
		\quad + same-paper graph augmentation                  & 0.227          & $\sim$140   \\
		\midrule
		Gemini 3.1 Pro                     & 0.170          & $\sim$60           \\
		GPT-5.5                            & 0.264          & $\sim$60        \\
		\bottomrule
	\end{tabular}
\end{table}

\begin{figure}[h]
	\centering
	\begin{minipage}{0.49\textwidth}
		\centering
		\includegraphics[width=\linewidth]{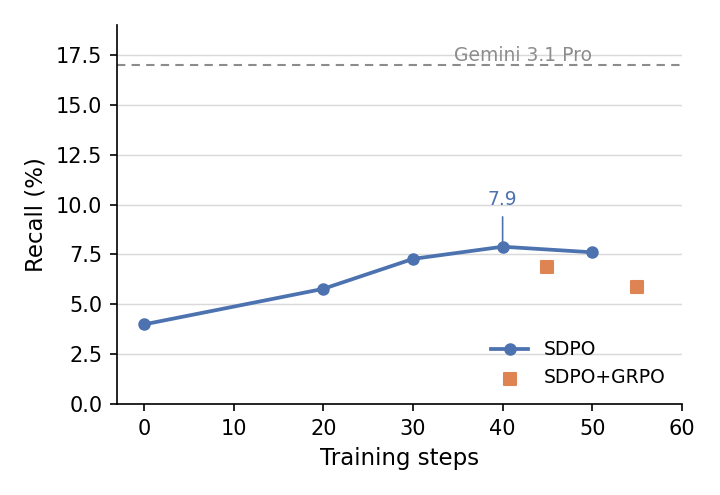}
		\caption{Validation recall ($k{=}10$, $H{=}6$) over SDPO training. Recall rises above the base model and then flattens near 8\%, below the prompted Gemini baseline.}
		\label{fig:traj}
	\end{minipage}\hfill
	\begin{minipage}{0.49\textwidth}
		\centering
		\includegraphics[width=\linewidth]{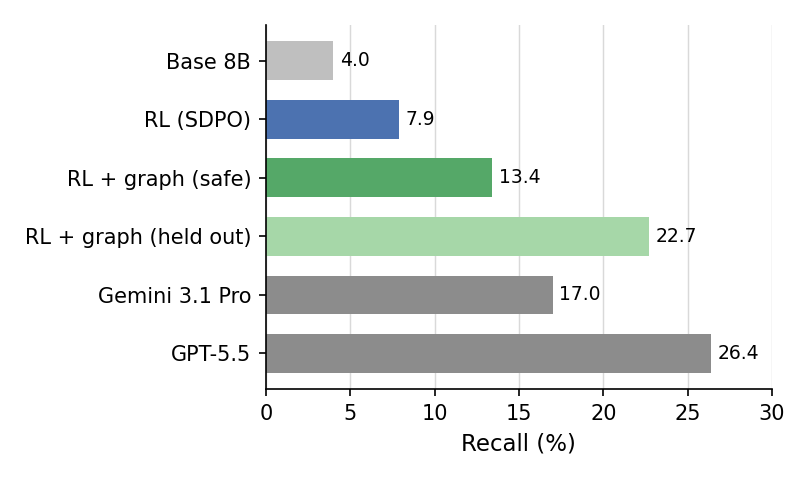}
		\caption{Recall at $k{=}10$. Cross-paper graph expansion roughly doubles the trained policy's recall.}
		\label{fig:systems}
	\end{minipage}
\end{figure}

\paragraph{Policy improvement saturation.} Figure~\ref{fig:traj} shows validation recall during SDPO training. The untrained Qwen3-8B policy retrieves, on average, 4\% of a target's true dependencies. Adding a short reasoning trace and training with SDPO raises recall to 7.9\%. After that point, further optimization did not meaningfully improve the result. A learning rate of $10^{-6}$ reaches 7.9\%, while $5{\times}10^{-6}$ reaches 7.7\%. Increasing the training set from 1{,}000 to 5{,}060 targets lowers recall to 5.8\%. These results place all trained policies in the same roughly 6--8\% band.

\paragraph{Citation-graph expansion recovers dependencies that embedding search misses.} Cross-paper graph expansion raises recall from 7.9\% to 13.4\% (Table~\ref{tab:main} and Figure~\ref{fig:systems}). This is because the policy often retrieves a statement near the target, and that nearby statement may cite some of the same background results as the target. One graph hop can therefore turn a semantic near miss into a true dependency. This change closes most of the gap to Gemini-3.1-Pro, although it also increases the candidate set from roughly 60 statements to roughly 115, so the comparison is not on an equal-budget basis. The same-paper expansion reaches 22.7\%, but this is not a valid result for premise selection. Theorems in the same paper share references and often have overlapping dependency lists. Allowing expansion through same-paper siblings therefore lets the method recover citations from the structure of the evaluation set rather than from a dependency-aware query strategy. For this reason, the cross-paper number is the main graph result.

\begin{figure}[h]
	\centering
	\includegraphics[width=0.5\linewidth]{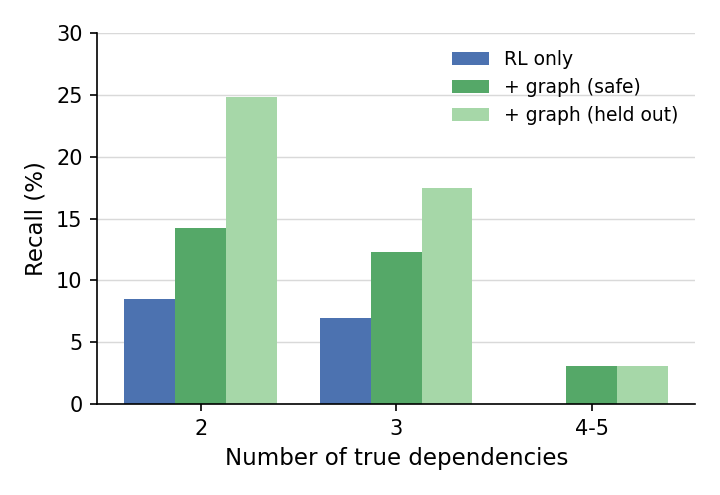}
	\caption{Recall at $k{=}10$ by number of true dependencies. Cross-paper graph expansion is most effective for low-degree targets.}
	\label{fig:buckets}
\end{figure}

\subsection{Discussion}

\paragraph{Query precision.} The prompted GPT-5.5 and Gemini-3.1-Pro baselines use the same embedding index and a comparable query budget, but retrieve more true dependencies than the trained Qwen3 policy. The trained policy also uses its budget and avoids repeated queries, so the main failure mode is not search control. Instead, it issues mathematically plausible topic queries that rarely name or characterize the exact earlier theorem cited by the proof. This also explains the negative ablations: more data provide little additional signal when most rollout groups contain no successful retrieval.

\paragraph{Graph expansion exposes useful structure.} Cross-paper expansion improves recall because a retrieved statement can be useful even when it is not itself a dependency: if it is close in the citation graph, its own dependencies may overlap with the target's. The gain is largest for low-degree targets, where a few good seeds can cover much of the dependency set (Figure~\ref{fig:buckets}). The same-paper diagnostic shows why this must be evaluated carefully. Same-paper expansion reaches 22.7\%, but that gain comes from shared bibliographies and local paper structure, so it is reported only as leakage.

\paragraph{Limitations.} Differences within the 6--8\% validation band should be treated cautiously because the trained-policy validation set has only $n{=}150$. However, cross-paper graph expansion adds another 5.5 points, which appears to be the more significant statistic. The graph result also uses a larger candidate set, increasing retrieval from roughly 60 to roughly 115 statements, so future systems should pair expansion with re-ranking.

\subsection{Conclusion}
Reinforcement learning improves a Qwen3-8B query policy for premise selection from roughly 4\% to 7.9\% dependency recall at $k{=}10$ under a six-query budget. The improvement then saturates across learning rates and training data scales. This suggests the limiting factor is query precision: the policy learns plausible mathematical searches, but it rarely identifies the exact prior theorem that the target proof cites. Cross-paper citation-graph expansion raises recall to 13.4\% by using retrieved statements as hints for dependency traversal. Future work should focus on learned graph expansion, re-ranking expanded candidates to control budget, and training policies that decide which retrieved statements are worth expanding.

\section{Node Distance and Embedding Distance}

We examine whether statements that are closer in the dependency graph also
have more similar slogan embeddings. Specifically, we measure the cosine
similarity between pairs of statements separated by a directed graph distance of $d$ dependency edges. Figure~\ref{fig:cos-dist} reports results for parent-only trajectories.

\subsection{Setup}

For each formality and depth $d$, we collected 1,000
accepted statement pairs. Starting from a randomly sampled initial statement seed, we repeatedly sampled one parent/dependency uniformly at random. At depth $d$, we retained the resulting pair only when its directed shortest-path distance from the seed was exactly $d$. Thus, accepted walks contain no directed shortcut of length less than $d$.

Each depth was sampled independently. In particular, failed walks were
replaced, and a new initial seed redrawn, until reaching the target number of accepted observations, so the
samples at depth $d$ should not be interpreted as continuations of the same
trajectories used at depth $d-1$.

For embedding comparison, we used sloganized statements embedded with
Qwen3-8B. Embeddings are $\ell_2$-normalized before cosine similarity is
computed as their dot product. Error bars show pointwise 95\%
normal-approximation confidence intervals for the mean cosine similarity.
The dashed horizontal line denotes a graph-free random-pair baseline,
computed separately for the formal and informal corpora as the mean cosine
similarity over 1,000 randomly paired, distinct statements.

\subsection{Results}

\begin{figure}[h]
    \centering
    
    \begin{subfigure}{0.48\textwidth}
        \centering
        \includegraphics[width=\textwidth]{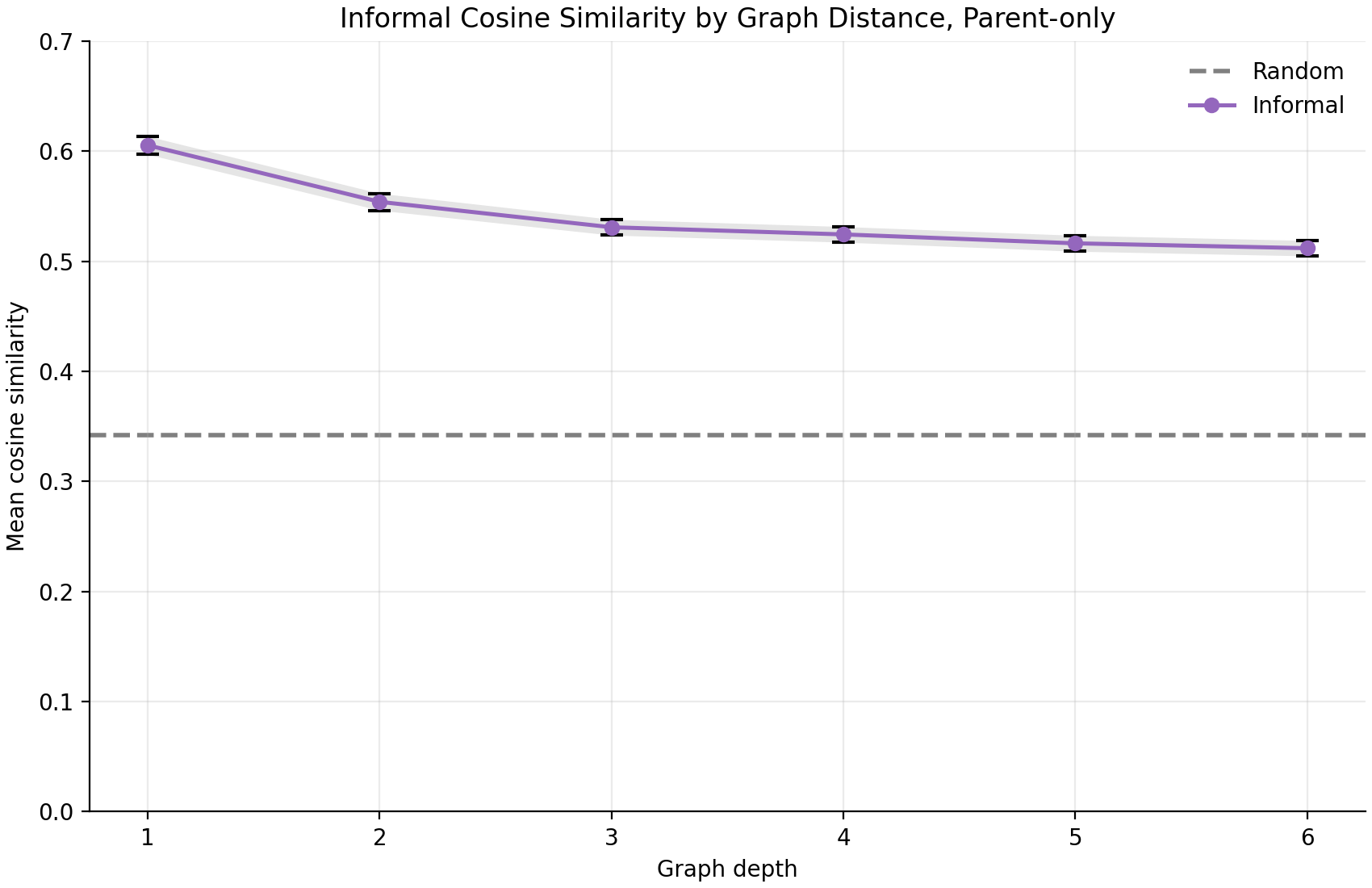}
        \caption{Informal cosine similarity by $d$-hop distance.}
        \label{fig:graph1}
    \end{subfigure}
    \hfill
    \begin{subfigure}{0.48\textwidth}
        \centering
        \includegraphics[width=\textwidth]{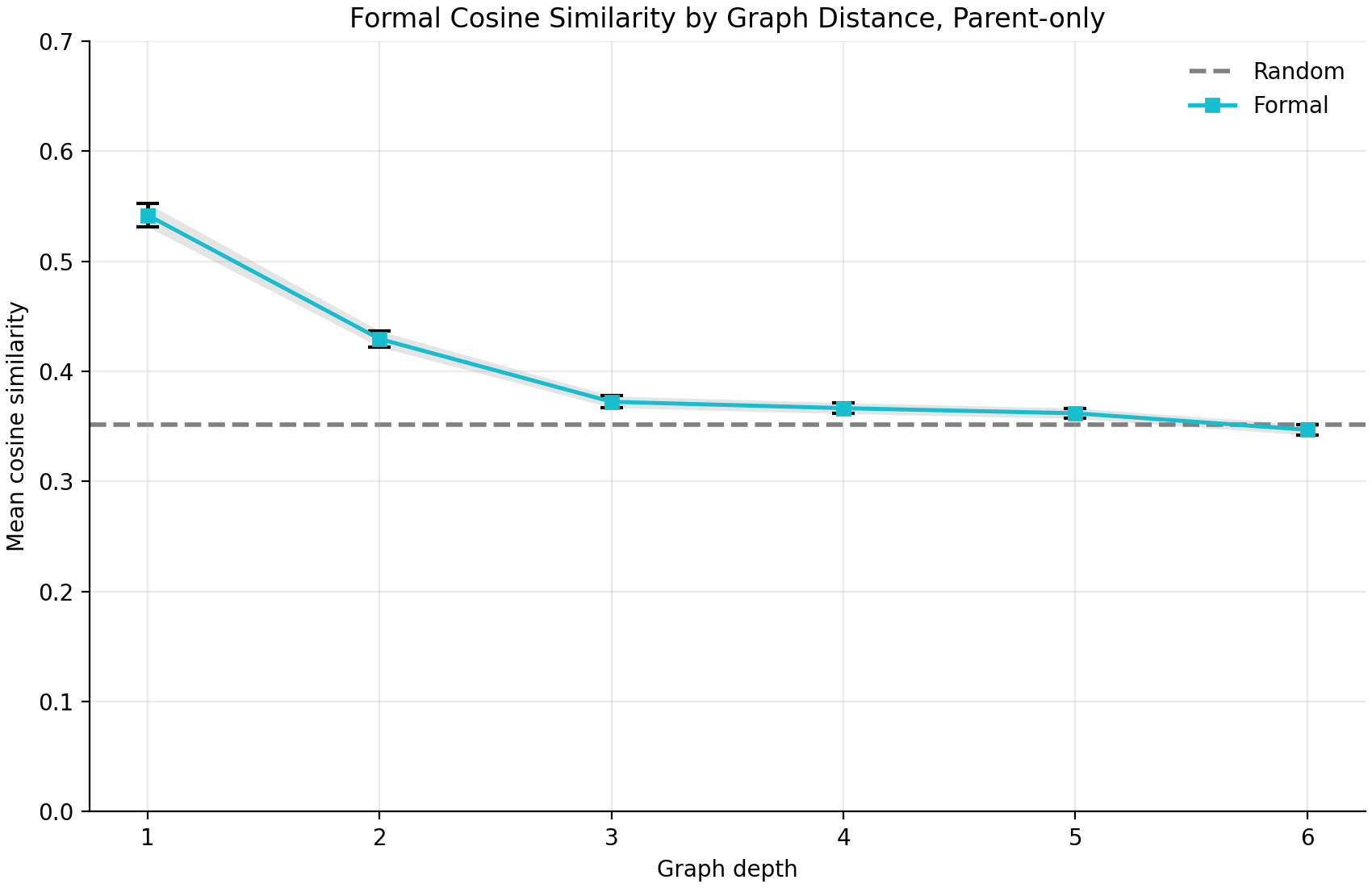}
        \caption{Formal cosine similarity by $d$-hop distance.}
        \label{fig:graph2}
    \end{subfigure}
    
    \caption{Cosine similarity and node depth for informal and formal graphs.}
    \label{fig:cos-dist}
\end{figure}

Figure~\ref{fig:cos-dist} shows an overall decline in cosine similarity as
directed parent-only graph distance increases in both graphs. The decline is substantially sharper in the formal graph: mean cosine similarity falls from approximately $0.54$ at depth one to roughly the random-pair baseline by depth six. In contrast, informal pairs remain highly similar even at larger graph distances, declining from approximately $0.61$ at depth one to about $0.51$ at depth six, well above the corresponding random baseline.

Thus, under parent-only dependency paths, informal statements retain greater embedding similarity over longer graph distances than formal statements. One possible explanation for the faster decline in the formal graph is its substantially higher local connectivity. Lean records fine-grained dependencies on definitions, typeclass instances, inherited structure, and other reusable mathematical infrastructure. Theorems about a narrow topic may have parent dependencies that connect it to broadly applicable declarations, such as general algebraic or module-theoretic results. Although these edges represent explicit and valid formal dependencies, they need not preserve strong high-level semantic similarity in the generated slogan space. By contrast, informal dependency paths may more often remain within a shared local development or exposition, allowing statements to retain greater embedding similarity over several graph steps.

\subsection{Limitations}

This analysis conditions on successful trajectories whose endpoints have usable slogan embeddings and whose directed shortest-path distance is exactly $d$. As depth increases, obtaining a fixed number of accepted pairs may therefore favor seed statements with sufficiently rich local neighborhoods, longer dependency chains, and reachable embedding-eligible endpoints. The resulting samples may not be representative of all statements at a given graph distance.

Moreover, each depth is sampled independently, and seed statements, nodes, and local neighborhoods may recur across sampled trajectories. The reported pointwise confidence intervals treat observations as independent and therefore may not fully capture uncertainty induced by overlap among walks.

\section{In-Context Literature Search}

Research mathematicians often need to identify a known result from a partial description embedded in an ongoing proof or exposition. We study a related task: recovering a cited theorem from a local passage of mathematical text after its citation has been removed.

\subsection{Methodology}

We evaluate whether language models can recover cited mathematical results from redacted arXiv passages using either unrestricted web search or retrieval over our mathematical statement database. We sampled 149 passages from mathematics arXiv papers released after February 2026. Restricting the sample to recent papers was intended to reduce the likelihood that the evaluated passages were available in the models' pretraining data.

Each selected passage contained a citation command referring to a named theorem, lemma, proposition, or analogous statement in another arXiv paper. We replaced the citation with \texttt{[Citation Needed]} and removed the corresponding bibliographic entry from the source paper's \texttt{.bib} file. Models received the same redacted passage and complete remaining bibliography in every condition.

We evaluate three retrieval settings. In the web-search baseline, models had unrestricted web access. In the dense-retrieval condition, models could issue repeated searches against our statement database, receiving the top 10 retrieved candidates for each query. In the graph-based condition, models could repeatedly search and navigate the statement dependency graph. An \emph{exact match} requires the model to return both the correct theorem identifier and the cited paper's arXiv identifier; a \emph{paper-level match} requires recovery of the correct cited arXiv paper, irrespective of the theorem identifier.

\subsection{Results}

\begin{table}[h]
\centering
\setlength{\tabcolsep}{4pt}
\small
\begin{tabular}{lrr}
\toprule
\textbf{System} & \textbf{Exact Match} & \textbf{Paper-Level Match} \\
\midrule
\multicolumn{3}{c}{\textbf{Web Search Baseline}} \\
ChatGPT 5.4       & 0.336 & 0.523 \\
Gemini 3.1 Pro    & \textbf{0.570} & \textbf{0.752} \\
Claude Sonnet 4.5 & 0.181 & 0.443 \\
\midrule
\multicolumn{3}{c}{\textbf{Dense Retrieval}} \\
ChatGPT 5.4       & 0.228 & 0.409 \\
Gemini 3.1 Pro    & 0.181 & 0.409 \\
Claude Sonnet 4.5 & 0.121 & 0.396 \\
\midrule
\multicolumn{3}{c}{\textbf{Graph-Based Retrieval}} \\
ChatGPT 5.4       & 0.255 & 0.456 \\
Gemini 3.1 Pro    & 0.289 & 0.597 \\
Claude Sonnet 4.5 & 0.148 & 0.322 \\
\bottomrule
\end{tabular}
\caption{
In-context citation recovery on 149 redacted arXiv passages. Exact match
requires the correct theorem identifier and cited paper arXiv identifier;
paper-level match requires only the correct cited paper.
}
\label{tab:literature-search}
\end{table}

Graph-based retrieval improves exact-match performance relative to dense
retrieval for all models, and improves paper-level
retrieval for ChatGPT 5.4 and Gemini 3.1 Pro. However, neither database-backed method outperforms unrestricted web search. The largest gap occurs for Gemini 3.1 Pro, which reaches a 0.570 exact-match rate with web search compared with 0.289 under graph-based retrieval.

These results suggest that graph navigation can provide useful additional structure beyond dense statement retrieval, but that the current retrieval interface and agent behavior do not yet match the effectiveness of unrestricted web search for this task.

\subsection{Limitations}

The web-search baseline is not a controlled retrieval comparison. Because the models can issue unrestricted searches, they may recover the original source paper by searching an exact quotation from the surrounding passage. Once that paper is identified, the redacted citation can often be recovered directly from the source document. Thus, web-search performance partly reflects access to the original evaluation instance rather than solely a model's ability to infer the missing citation from mathematical context.

The experiment also permits repeated retrieval and navigation actions without a matched search budget across conditions. Performance therefore reflects a combination of retrieval quality, agent search strategy, stopping behavior, and the practical affordances of each interface. In particular, dense retrieval returns a fixed top-10 candidate set per query, whereas graph-based retrieval permits open-ended local navigation.

Our evaluation set is limited to 149 examples and contains only citations that explicitly name a theorem-like result in an arXiv paper. It therefore does not represent the broader range of scholarly citations, including citations for background, methods, attribution, or entire lines of work. Moreover, the ground-truth citation may not be the only mathematically appropriate response: a model can identify a related theorem or an alternative source that supports the passage while still being scored as incorrect.

\end{document}